\documentclass[aps,prc,twocolumn,superscriptaddress,nofootinbib,preprintnumbers]{revtex4-2}

\usepackage{xcolor}
\usepackage{amsmath}
\usepackage{amssymb}
\usepackage{graphicx}
\usepackage{multirow}
\begin{document}
\preprint{INT-PUB-24-032}
%\preprint{}

\title{Bulk Viscosity in Dense Nuclear Matter}

% repeat the \author \affiliation  etc. as needed


\author{Steven P.~Harris}
%\email[]
\affiliation{Institute for Nuclear Theory, University of Washington, Seattle, WA 98195, USA}
\altaffiliation{Present address: Center for the Exploration of Energy and Matter and Department of Physics,
Indiana University, Bloomington, IN 47405, USA}

\date{July 27, 2023}
\begin{abstract}
Neutron star mergers are an extreme environment where, within milliseconds, the cold, dense matter in neutron stars is dramatically heated to some of the hottest temperatures encountered in astrophysics.  Matter in neutron star mergers is located solidly in the interior of the QCD phase diagram, and therefore the observation and simulation of these mergers provides an opportunity to learn about the strong interaction at a set of densities and temperatures that compliments those obtained in relativistic heavy-ion collisions.  Matter during the merger of two neutron stars is far from hydrostatic equilibrium, allowing for the study of not only the equation of state of dense matter, but of transport properties as well.  Perhaps the most likely transport property to play a role in neutron star mergers is the bulk viscosity, which acts to resist changes in density.  The bulk viscosity of matter at high densities has long been anticipated to damp oscillations in isolated neutron stars, and the study of its role in neutron star mergers is just beginning.

In this chapter, I describe bulk viscosity as a general concept, and then focus on bulk viscosity in the dense matter present in compact objects.  While this review is focused on bulk viscosity in the conditions present in neutron star mergers, I present a history of bulk viscosity research in dense matter, from its role in damping radial oscillations in neutron stars through its current applications in neutron star mergers.  The majority of the chapter consists of calculations of the bulk viscosity from Urca processes in generic neutron-proton-electron ($npe$) matter, and then in dense matter containing muons ($npe\mu$ matter) as well.  I make several approximations in these calculations to keep the focus on the concepts.  More precise calculations exist in the literature, to which I refer the reader.  One concept I attempt to elucidate is the thermodynamic behavior of a fluid element throughout an oscillation and how that leads to bulk-viscous dissipation.  I conclude with a discussion of the recent research into the role of weak interactions and bulk viscosity in neutron star mergers. 
\end{abstract}
\maketitle
\section{Bulk viscosity and transport in dense matter}\label{intro}

The nature of matter at the highest densities is a question of fundamental interest.  This matter can be studied in the laboratory with heavy-ion collisions, but also in astrophysical environments.  Neutron stars, first theoretically proposed in the 1930s \cite{Yakovlev:2012rd}, and observed as pulsars in the 1960s \cite{Schaffner-Bielich:2020psc}, contain matter of densities up to several times nuclear saturation density ($n_0\equiv 0.16\text{ fm}^{-3}$).  While the matter in isolated neutron stars is cold, in the sense that the temperature is much less than the Fermi energies of the constituent particles (termed \textit{degenerate}), when two neutron stars merge, shocks heat up the neutron star matter to temperatures of tens of MeV\footnote{In this chapter, I use natural units, where $c=\hbar=k_B=1$.  In these units, $1\text{ MeV} \approx 1.2\times 10^{10} \text{ K}$.}.  Neutron stars and neutron star mergers, together with heavy-ion collisions and nuclear experiments, probe complimentary regions of the QCD phase diagram \cite{Dexheimer:2020zzs}.

\subsection{Advantages of studying transport}
It is most common to study the equation of state (EoS) of dense matter, a relationship between its energy density $\varepsilon$ and the pressure $P$.  This is important in its own right, but also because it leads to the prediction of masses and radii of neutron stars (through the solution of the Tolman-Oppenheimer-Volkoff (TOV) equations), the tidal deformability of a neutron star, and the oscillation frequencies of isolated neutron stars and differentially rotating neutron star merger remnants.  However, the EoS $\varepsilon=\varepsilon(P)$ does not tell us the full nature of dense matter, because there is no unique relationship between the degrees of freedom of the dense matter and the EoS.  This ``masquerade problem'' implies that a hybrid star (a neutron star with a quark core) can have a similar mass and radius as a neutron star containing no quark matter \cite{Alford:2004pf}.  Therefore, it is important to study other aspects of dense matter beyond the EoS, including transport properties.  In condensed matter physics, one might study the electrical conductivity, thermal conductivity, or the heat capacity of a sample, which would provide insight into the low-energy degrees of freedom around the Fermi surface \cite{2011cmn..book.....M}.  In dense matter, transport properties like the thermal conductivity, specific heat, and both shear and bulk viscosity have been calculated and, if deemed to be significant, implemented in simulations of neutron star cooling \cite{Yakovlev:2004iq}, core-collapse supernovae \cite{Mezzacappa:2020oyq}, and neutron star mergers (Ref.~\cite{Foucart:2022bth} and Sec.~\ref{sec:bv_effects} of this chapter). 
 A fairly recent review article by Schmitt \& Shternin \cite{Schmitt:2017efp} provides a comprehensive summary of transport in cold, dense matter.
%%%%%%%%%%%%%%%%%%%%%%%%%%%%%%%%%%%%%%%%%%%%%%%%%%%
\subsection{Bulk viscosity in all types of fluids}
In this chapter, I discuss the bulk viscosity of dense matter.  Bulk viscosity arises in a fluid that experiences a change in its density, causing some internal degree of freedom within the matter to be pushed out of equilibrium.  In response, internal processes ``turn on'' in order to restore equilibrium.  In the chemistry literature, this negative-feedback effect is known as Le Chatelier's principle \cite{Glendenning:1997wn}.  Throughout the duration of the expansion or contraction of the system, the (irreversible) internal processes that attempt to restore equilibrium dissipate energy from the motion of the matter and turn it into heat.  The amount of bulk-viscous dissipation depends on the relaxation timescale relative to the timescale of the density change.  If the relaxation timescale is very short, then the system is hardly out of equilibrium at all, and the bulk viscosity is small.  If the relaxation timescale is very long, then it is as if no internal relaxation occurs, and the bulk viscosity is also small.  As the two timescales become comparable, bulk viscosity grows larger and could be significant.  A textbook discussion of bulk viscosity in fluids is given in Landau \& Lifshitz \cite{1987flme.book.....L}.

Bulk viscosity exists in many types of fluids, from gases encountered in everyday life to fluids at the most extreme densities and temperatures in nature.  For example, in diatomic gases, a density change leads to a change in kinetic energy of the molecules, which at some finite rate, turns into vibrational and rotational energy \cite{1991mtnu.book.....C}.  In cosmology, an expanding mixture of matter and radiation are pushed out of thermal equilibrium and exchange energy at some finite rate, leading to energy dissipation \cite{Weinberg:1972kfs,Zimdahl:1996fj}.  In the expanding quark-gluon plasma formed in relativistic heavy-ion collisions, the matter departs from thermal equilibrium and is reequilibrated by strong interactions \cite{Arnold:2006fz}.  A neutron Fermi liquid, devoid of flavor-changing interactions, has a bulk viscosity stemming from binary collisions which attempt to restore thermal equilibrium \cite{1970AnPhy..56....1S,baym2008landau,Kolomeitsev:2014gfa}. 

In neutron stars, as the rest of the chapter will demonstrate, a density change pushes the matter out of chemical equilibrium, and flavor-changing interactions (for example, Urca processes) turn on to restore equilibrium, generating entropy as they proceed.  A fluid element in the neutron star undergoing a small-amplitude sinusoidal density oscillation will traverse a path in the pressure-volume ($PV$) plane.  If the chemical equilibration occurs very fast or very slow compared to the oscillation frequency, the $PV$ curve will be a line that the fluid element follows forward and then backward over the course of the oscillation.  If the chemical equilibration occurs at a similar timescale as the density oscillation, the fluid element will follow a closed curve in the $PV$ plane.  The area enclosed by that curve indicates the amount of work done on the fluid element, and is maximal when the rate of equilibration matches the density oscillation frequency, a resonance-like behavior.  This mechanism of energy dissipation will be explored in much greater detail in the rest of the chapter.
%%%%%%%%%%%%%%%%%%%%%%%%%%%%%%%%%%
\subsection{History of bulk viscosity in dense matter} \label{sec:history}
%%%%%%%%%%%%%%%%%%%%%%%%%%%%%%%%%%%%
\subsubsection{The early years: 1965 - 1980} \label{sec:history_early}
The study of bulk viscosity in neutron stars started in the mid-1960s, following a resurgence of neutron star research in the late 1950s and early 1960s, where midcentury developments in nuclear interactions and hypernuclear physics were first applied to neutron star structure calculations. 
 For broader context, detailed histories of nuclear physics and its application in neutron stars are presented in the textbooks \cite{Schaffner-Bielich:2020psc,Glendenning:1997wn,Haensel:2007yy,Shapiro:1983du}.  In the mid-1960s, researchers were interested in the radial oscillation modes of neutron stars presumed to arise in their initial formation in supernovae.  Curiosity about the duration of these oscillations after the birth of the neutron star, as well as the possibility of the conversion of vibrational energy into heat, leading to enhanced radiation from the neutron star surface (see the contemporary review articles \cite{Wheeler:1966tg,Cameron:1970wm}), lead to preliminary calculations (Finzi \cite{1965PhRvL..15..599F}, Meltzer \& Thorne \cite{1966ApJ...145..514M}, and Hansen \& Tsuruta \cite{1967CaJPh..45.2823H}) of vibrational energy dissipation arising from chemical equilibration.  In these early works, which considered neutron stars built of neutron-proton-electron ($npe$) matter (see Sec.~\ref{sec:npe_bv}), modified Urca processes (Sec.~\ref{sec:npe_bv_and_beq}) were the equilibration mechanism.
 
 The calculations of Finzi \& Wolf \cite{1968ApJ...153..835F} and Langer \& Cameron \cite{1969Ap&SS...5..213L} in the late '60s set the stage for almost all subsequent neutron star bulk viscosity work.  Finzi \& Wolf calculated the time evolution of both the temperature and radial pulsation amplitude of a model neutron star, accounting for the chemical equilibration (via modified Urca) throughout a cycle of the oscillation, as well as the associated increase in the neutrino emissivity.  They calculated the $PdV$ work done on the fluid element through the oscillation.  Langer \& Cameron considered hyperonic matter undergoing a sinusoidal density oscillation and strangeness-changing interactions that attempt to reequilibrate the strangeness.  This work was the first to spell out the resonant structure of this type of energy dissipation (see Sec.~\ref{sec:npe_bv_and_beq}).  Jones was the first to attribute the vibrational damping to a hydrodynamic bulk viscosity coefficient, in his study of hyperonic reactions in pulsating dense matter \cite{1970ApL.....5...33J}.  

The 1970s were very quiet in terms of neutron star bulk viscosity research.  However, during that decade the understanding of neutrino-nucleon interactions expanded dramatically, especially with the experimental confirmation of the neutral current interaction in 1973, and it became clear that neutrinos are trapped in core-collapse supernovae and therefore neutrino transport schemes \cite{Lattimer:1981hn} would have to be developed\footnote{At present, it seems most likely that the implementation of bulk-viscous effects in neutron star merger simulations will be done through the neutrino transport scheme (see Sec.~\ref{sec:bv_effects}).}.  After a calculation of neutrino mean free paths (MFPs) in dense matter \cite{Sawyer:1978qe}, in a 1980 paper \cite{Sawyer:1980wp} Sawyer calculated the bulk viscosity of proto-neutron star matter where neutrinos are trapped.
%%%%%%%%%%%%%%%%%%%%%%%%%%%%%%%%%%%%%%%%%%%%%%
\subsubsection{Expansion of applications in cold dense matter: 1980-2000}\label{sec:history_middle}
The 1980s saw the first calculations of bulk viscosity in (non-interacting) quark matter, by Wang \& Lu \cite{1985AcApS...5...59W,Wang:1984edr}, and later, Sawyer \cite{Sawyer:1989uy}.  Up until the end of the 1980s, essentially all dense-matter bulk viscosity calculations had in mind a neutron star undergoing radial oscillations that would be damped by bulk viscosity.  In 1990, Cutler, Lindblom, \& Splinter \cite{1990ApJ...363..603C}, armed with an improved calculation of bulk viscosity in $npe$ matter by Sawyer \cite{Sawyer:1989dp}, calculated the contribution of the bulk viscosity to the damping time of nonradial oscillations of cold neutron stars, and delineated the temperature regimes in which shear and bulk viscosity are each the dominant microscopic dissipation mechanisms (this switchover between the two dissipation mechanisms became more famous later in the context of the r-mode instability window).

In a 1991 paper, Lattimer \textit{et al.}~\cite{Lattimer:1991ib} pointed out that the direct Urca process, long thought (except by Boguta \cite{Boguta:1981mw}) to be forbidden in cold neutron stars because they were too neutron-rich, may occur in the densest regions of a neutron star, where the proton fraction $x_p\equiv n_p/n_B \gtrsim 0.11-0.15$.  This result upended prior understandings of neutron star cooling, where it had been thought that if cooling significantly faster than predicted by modified Urca was observed, it was likely due to an exotic phase of matter, like quark matter or a pion condensate.  In cold systems, where the beta equilibration rate is much slower than the timescale of the density change, the bulk viscosity is proportional to the beta equilibration rate (see Eq.~\ref{eq:zeta_cold}), and therefore is strongly enhanced when direct Urca is present (Haensel \& Schaeffer \cite{Haensel:1992zz}).

In 1992, Madsen \cite{Madsen:1992sx} studied the bulk viscosity in strange quark matter, extending it to the \textit{suprathermal} regime\footnote{Most bulk viscosity calculations up to this point considered \textit{subthermal} density oscillations, where the matter is pushed only slightly out of chemical equilibrium $\delta\mu\ll T$. 
 \textit{Suprathermal} oscillations push the matter far out of chemical equilibrium $\delta\mu\gg T$, though the formalism developed in this chapter is only sensible when the amplitude of the density oscillation remains small.  The first use of the terms \textit{subthermal} and \textit{suprathermal} I could find is by Haensel, Levenfish, \& Yakovlev \cite{Haensel:2002qw}.}.  Bulk viscosity in the suprathermal regime is enhanced by potentially orders of magnitude compared to the subthermal case.  Madsen's suprathermal bulk viscosity work was extended by Goyal \textit{et al.}~\cite{Goyal:1993ch} and then by Gupta \textit{et al.}~\cite{Gupta:1997ce}, who calculated the suprathermal bulk viscosity of nuclear matter equilibrating via direct Urca.  The calculation of Gupta \textit{et al.}~went essentially unnoticed in the literature, and was recalculated by Alford, Mahmoodifar, \& Schwenzer \cite{Alford:2010gw}, who account for the more likely possibility of $npe$ matter equilibrating via modified Urca.

Another important development in the 1990s was the start of a movement away from the bulk viscosity coefficient, and back to tracking the chemical reactions themselves, at least, in certain situations.  Gourgoulhon \& Haensel \cite{1993A&A...271..187G} studied the the collapse of a neutron star to a black hole, and the ensuing weak interactions that occur due to the compression of the dense matter.  Here it is clearly impossible to use the small-amplitude, oscillation-averaged bulk viscosity formalism discussed in this chapter.  This study was extended to neutron stars near their maximum allowed mass, where the particle content was evolved along with the hydrodynamic equations during a stellar oscillation \cite{1995A&A...294..747G}.  Another physical situation where chemical reactions themselves were tracked is rotochemical heating, studied by Reisenegger \cite{Reisenegger:1994be}.  As a neutron star spins down, the centrifugal force decreases and the star contracts, pushing the matter out of chemical equilibrium.  The ensuing chemical reactions heat the matter (though, accounting for neutrino cooling, there may or may not be \textit{net} heating).  Any temperature change feeds back on the equilibration rates, and the deviation from chemical equilibrium $\delta\mu$ and the temperature $T$ evolve in nontrivial ways.  This analysis was extended to strange stars by Cheng \& Dai \cite{Cheng:1996it}.  

At the end of the 1990s, Andersson \cite{Andersson:1997xt} and Friedman \& Morsink \cite{Friedman:1997uh} conducted the first general-relativistic study of the r-mode, a nonradial mode present in rotating stars.  They found that it (in the absence of viscosity) is unstable for all rotation rates, a consequence of the CFS instability \cite{Andersson:2019yve}.  As viscosity can stabilize the r-mode, Andersson, Kokkotas, \& Schutz \cite{Andersson:1998ze} used shear and bulk viscosity calculations to map out the spin rates and core temperatures for which the r-mode is unstable (the \textit{r-mode instability window}).  The r-mode and instability window literature is so vast, that I will not attempt to detail it here (Andersson's textbook \cite{Andersson:2019yve} is a nice reference), but after these calculations, the r-mode became a favorite context in which to study bulk viscosity.
%%%%%%%%%%%%%%%%%%%%%%%%%%%%%%%%
\subsubsection{Bulk viscosity pre-GW170817: 2000-2017}\label{sec:history_before_gw170817}
Until about 2000, even though most bulk viscosity calculations were done with cold neutron star matter in mind, no calculation had included the nucleon Cooper pairing that is expected to occur for temperatures less than a few times $10^9$ K.  This \textit{critical temperature} is a function of density and is still largely uncertain today \cite{Sedrakian:2018ydt}.  Pairing produces a gap in the excitation spectrum at the Fermi surface, exponentially suppressing the rate of beta equilibration in degenerate matter \cite{Yakovlev:1999sk,Yakovlev:2000jp}.  In a series of papers in the early 2000s, Haensel, Levenfish, \& Yakovlev calculated the bulk viscosity in superfluid nuclear matter due to the direct Urca process \cite{Haensel:2000vz}, the modified Urca process \cite{Haensel:2001mw}, and in hyperonic matter \cite{Haensel:2001em}.  Because at low temperatures the subthermal bulk viscosity is proportional to the beta equilibration rate (Eq.~\ref{eq:zeta_cold}), it is strongly suppressed by superfluidity.  However, this suppression can be overcome in the suprathermal regime due to ``gap-bridging'' \cite{Alford:2011df,Alford:2016cee}.  Finally, Gusakov \cite{Gusakov:2007px} pointed out that in superfluid hydrodynamics, there are actually multiple bulk viscosity coefficients, not just one, and they all contribute significantly to the energy dissipation in superfluid nuclear matter.  Gusakov \& Kantor \cite{Gusakov:2008hv} extended this analysis to matter with superfluid nucleons and hyperons.  

The next decade saw a proliferation of bulk viscosity calculations in ever more exotic forms of matter.  Bulk viscosity in hyperonic matter was revisited by Jones \cite{Jones:2001ya}, Lindblom \& Owen \cite{Lindblom:2001hd}, van Dalen \& Dieperink \cite{vanDalen:2003uy}, and Chatterjee \& Bandyopadhyay \cite{Chatterjee:2006hy}.  Bulk viscosity in quark/hadron mixed phases was calculated by Drago, Lavagno, \& Pagliara \cite{Drago:2003wg} and Pan, Zheng, \& Li \cite{Pan:2006dp}.  A myriad of quark matter phases were studied, including unpaired phases of strange quark matter by Sa'd, Shovkovy, \& Rischke \cite{Sad:2007afd}, Alford, Mahmoodifar, \& Schwenzer \cite{Alford:2010gw} in the suprathermal regime, and Shovkovy \& Wang \cite{Shovkovy:2010xk} for anharmonic density oscillations.  The bulk viscosity for paired quark matter phases like 2SC (Alford \& Schmitt \cite{Alford:2006gy}) and other phases that have at least one unpaired quark flavor (Sa'd, Shovkovy, \& Rischke \cite{Sad:2006egl}, Wang \& Shovkovy \cite{Wang:2010ydb} and Berdermann \textit{et al.}~\cite{Berdermann:2016mwt}), as well as the color-flavor-locked (CFL) phase (Alford \textit{et al.}~\cite{Alford:2007rw}, Manuel \& Llanes-Estrada \cite{Manuel:2007pz}, and Alford, Braby, \& Schmitt \cite{Alford:2008pb}) were calculated.  Mannarelli \& Manuel \cite{Mannarelli:2009ia} and Bierkandt \& Manuel \cite{Bierkandt:2011zp} calculated all three of the bulk viscosity coefficients (c.f.~Gusakov \cite{Gusakov:2007px}) present in superfluid CFL quark matter.  Chatterjee \& Bandyopadhyay \cite{Chatterjee:2007qs} calculated the bulk viscosity of nuclear matter with a kaon condensate. 
Bulk viscosity from muon-electron conversion, potentially dominant in superfluid nuclear matter, was calculated by Alford \& Good \cite{Alford:2010jf}.  Finally, Huang \textit{et al.}~\cite{Huang:2009ue} studied viscosity in strange quark matter under a strong magnetic field, where there are now two bulk viscosity coefficients, one parallel and one perpendicular to the local magnetic field direction.  Out of these papers were several \cite{Sad:2007afd,Shovkovy:2010xk,Alford:2006gy,Wang:2010ydb,Dong:2007mb,Jaikumar:2010gf} that considered multiple equilibration channels, where the bulk-viscous behavior can become quite intricate.  Dense $npe\mu$ matter has multiple equilibration channels, and is explored in Sec.~\ref{sec:npemu_bv}.

A few papers considered the effect of chemical reactions on both dynamical and thermal evolution of neutron stars.  Rotochemical heating was extended to a general-relativistic formalism by Fernandez \& Reisenegger \cite{Fernandez:2005cg} and to superfluid nuclear matter by Petrovich \& Reisenegger \cite{Petrovich:2009yh}.  Gusakov, Yakovlev, \& Gnedin \cite{Gusakov:2005dz} evolved an oscillating neutron star in general relativity, tracking the oscillation energy dissipation due to bulk viscosity but also the heating from the chemical reactions, essentially expanding on the Finzi \& Wolf \cite{1968ApJ...153..835F} calculation.  Kantor \& Gusakov \cite{Kantor:2011yj} and Gusakov \textit{et al.}~\cite{Gusakov:2012zx} did similar analyses, but in superfluid neutron stars.     
%%%%%%%%%%%%%%%%%%%%%%%%%%%%%5
\subsubsection{Bulk viscosity after GW170817}\label{sec:history_after_gw170817}
In August 2017, the LIGO-VIRGO collaboration detected a gravitational wave signal \cite{LIGOScientific:2017vwq} from the inspiral of two merging neutron stars.  Just under two seconds later, a gamma-ray burst was detected by the Fermi Gamma-ray Burst Monitor and soon after, an optical counterpart was identified \cite{LIGOScientific:2017ync}, starting a new chapter in multimessenger astrophysics.  

Just prior to this event, Alford \textit{et al.}~\cite{Alford:2017rxf} considered transport properties in nuclear matter under conditions expected to be obtained in neutron star mergers, where the cold matter in the original neutron stars ($T \lesssim 10 \text{ keV}$ \cite{Arras:2018fxj}) is heated to temperatures of tens of MeV.  Alford \textit{et al.}~determined that bulk viscosity due to Urca processes might be relevant for neutron star mergers\footnote{The possibility of bulk viscosity in neutron star mergers was first considered many years earlier, in the inspiral stage by Lai \cite{Lai:1993di} and in the postmerger stage by Ruffert \& Janka \cite{Ruffert:2001gf}, but was judged not to be of primary importance.}, because the timescale on which a sizeable fraction of the energy of an oscillation can be dissipated was found to be milliseconds, well within the (estimated) lifetime of many neutron star merger remnants (see Fig.~3 in the review article \cite{Bernuzzi:2020tgt}).  The 2017 neutron star merger and the Alford \textit{et al.}~bulk viscosity analysis initiated new interest in weak interactions at high temperatures (say, $T\gtrsim 1\text{ MeV}$) and the bulk viscosity stemming from these weak interactions.  With only a few exceptions\footnote{For example, Yakovlev, Gusakov, \& Haensel \cite{Yakovlev:2018jia} calculated the bulk viscosity in nuclear pasta phases which, unexpectedly, can equilibrate via direct Urca (though at a suppressed rate) and Ofengeim \textit{et al.}~\cite{Ofengeim:2019fjy} improved past calculations of hyperonic bulk viscosity.}, most bulk viscosity papers written since 2017 have had neutron star mergers in mind.

To study viscous effects in the neutron star inspiral, Arras \& Weinberg \cite{Arras:2018fxj} considered a neutron star undergoing forced oscillations due to its companion neutron star, and found that suprathermal bulk viscosity from Urca processes heats the star up to, at most, tens of keV.  Ghosh, Pradhan, \& Chatterjee studied the same phenomenon but considering hyperonic interactions \cite{Ghosh:2023vrx}.  Most \textit{et al.}~\cite{Most:2021zvc} studied the effect of bulk viscosity during tidal deformation in the inspiral.  This analysis was refined recently by Ripley, Hegade, \& Yunes \cite{Ripley:2023qxo} and a new coefficient, the dissipative tidal deformability, was introduced.  Viscous effects in the inspiral phase are discussed further in Sec.~\ref{sec:inspiral}.

Alford \& Harris \cite{Alford:2018lhf} calculated the direct Urca rates in $T = 0.5-10 \text{ MeV}$ neutrino-transparent nuclear matter, going beyond the Fermi surface (FS) approximation (or strongly-degenerate limit) by doing the full integration over the phase space, and found that for $T\gtrsim 1\text{ MeV}$, the direct Urca rates become significant even below the direct Urca threshold density and furthermore, the traditional beta equilibrium condition in neutrino-transparent matter has to be modified.  Improvements to this calculation were later made by Alford \textit{et al.}~\cite{Alford:2021ogv}.  

Improving upon the back-of-the-envelope estimate of Alford \textit{et al.}~\cite{Alford:2017rxf}, Alford \& Harris \cite{Alford:2019qtm} used the improved direct Urca rates and corrected beta equilibrium condition in a calculation of the bulk viscosity in neutrino-transparent nuclear matter and showed that the bulk viscosity could damp 1 kHz density oscillations in certain thermodynamic conditions in as little as 5 ms, making the case that bulk viscosity should be included in numerical simulations of neutron star mergers\footnote{Very recently, Alford, Haber, \& Zhang \cite{Alford:2023gxq} found a correction of this calculation due to the density dependence of the difference between the zero-temperature and finite-temperature neutrino-transparent beta equilibrium conditions, and recalculated the bulk viscosity.}.  Alford, Harutyunyan, \& Sedrakian \cite{Alford:2019kdw,Alford:2020lla} calculated the bulk viscosity in nuclear matter hot enough to trap neutrinos ($T\gtrsim 5-10\text{ MeV}$ \cite{Alford:2018lhf}), and found that it is likely too small to have an effect on mergers, but also found interesting features in the bulk viscosity, such as conformal points.  Alford, Harutyunyan, \& Sedrakian \cite{Alford:2021lpp,Alford:2022ufz,Alford:2023uih} considered neutrino-trapped and neutrino-transparent nuclear matter with muons ($npe\mu$ matter) and calculated its bulk viscosity, including the reactions of the muons as well.

Alford \& Haber \cite{Alford:2020pld} recalculated the hyperonic bulk viscosity, trying to explore the conditions where merger temperatures are high enough to produce a thermal hyperon population, but found that the hyperon bulk viscosity in merger conditions is small.  Instead, the peak bulk viscosity was found to be at keV temperatures more relevant to the inspiral phase.  

In parallel with the microphysical calculations of the bulk viscosity, a reconsideration of bulk viscosity and chemical reactions within hydrodynamic frameworks has been ongoing.  Gavassino, Antonelli, \& Haskell \cite{Gavassino:2020kwo} confirmed, for a system like $npe$ matter with one independent particle fraction, the equivalence (for small deviations from equilibrium) between tracking the change of chemical composition as it evolves due to weak interactions and a Muller-Israel-Stewart (MIS) theory where a bulk stress $\Pi$ is introduced into the stress-energy tensor and evolves according to the MIS equation.  Gavassino \& Noronha \cite{Gavassino:2023xkt} recently generalized this equivalence to large deviations from equilibrium.  Gavassino \cite{Gavassino:2023eoz} showed that in a two-component system like $npe\mu$ matter, the near-equilibrium viscous dynamics are described by a bulk stress $\Pi$ that evolves in time not according to the MIS equation, but according to a Burgers equation.  In two papers, Camelio \textit{et al.}~compared the ``chemical composition tracking'' approach to the MIS approach by implementing both in a simulation of a radially oscillating neutron star \cite{Camelio:2022ljs,Camelio:2022fds}.

Finally, in the past couple years, bulk viscosity is beginning to be implemented in neutron star merger simulations.  Most \textit{et al.}~\cite{Most:2021zvc} postprocessed an inviscid merger simulation to predict the possible strength of bulk viscosity in the merger remnant.  Hammond, Hawke, \& Andersson \cite{Hammond:2021vtv} and Celora \textit{et al.}~\cite{Celora:2022nbp} contemplated the issues associated with implementation of bulk viscosity in merger simulations.  Hammond, Hawke, \& Andersson \cite{Hammond:2022uua} ran merger simulations with infinitely fast and infinitely slow Urca reactions.  A few very recent simulations have included flavor-changing interactions, including Radice \textit{et al.}~\cite{Radice:2021jtw}, Zappa \textit{et al.}~\cite{Zappa:2022rpd},  and Most \textit{et al.}~\cite{Most:2022yhe}, and just prior to submission of this chapter, Chabanov \& Rezzolla \cite{Chabanov:2023blf} put bulk viscosity into a merger simulation using the MIS formalism.  Bulk viscosity in merger simulations is discussed in Sec.~\ref{sec:postmerger}.
%%%%%%%%%%%%%%%%%%%%%%%%%%%%%%%%%%%%%%%%%%%%%%%%%%%
\section{Bulk viscosity in neutron-proton-electron matter}\label{sec:npe_bv}
%%%%%%%%%%%%%%%%%%%%%%%%%%%%%%%%%%%%%%%%
\subsection{Dense matter}
The nature of dense matter is not fully known.  In the vicinity of nuclear saturation density, the matter is uniform, with neutrons, protons, and electrons as degrees of freedom.  At low enough temperature, certainly below $T=1\text{ MeV}$, the nucleons form Cooper pairs and the matter becomes a superfluid \cite{Sedrakian:2018ydt}.  In a neutron star merger, after the stars collide, the vast majority of the matter rises to temperatures above 1 MeV, and is therefore not superfluid (though quark matter critical temperatures are expected to be higher \cite{Alford:2007xm}).  As the density increases significantly beyond $n_0$, new degrees of freedom may enter the system, including hyperons, a pion condensate, or the nuclear matter might undergo a phase transition to quark matter \cite{Alford:2019oge}.  As temperature is increased, thermal populations of particles, for example, pions \cite{Fore:2019wib}, may appear as well.

A sensible first step for modeling dense matter is to consider uniform $npe$ matter.  The electrons can be treated as a free Fermi gas, but strong interactions between the nucleons must be considered.  This is usually done with Skyrme energy density functionals \cite{Dutra:2012mb} or relativistic mean field theories (RMFs) \cite{Dutra:2014qga}.  I will use the IUF RMF \cite{Fattoyev:2010mx} throughout this chapter.  Of relevance to the discussion in this chapter, the IUF EoS (when muons are not added) has a direct Urca threshold near $n_B = 4n_0$ \cite{Alford:2021ogv}.  RMFs are motivated by the meson-exchange model of the strong interaction, and nucleons interact by exchanging sigma, omega, and rho mesons.  The mean field approximation is then taken, the meson dynamics are frozen, and the nucleons act like free particles but with effective masses and chemical potentials \cite{Glendenning:1997wn}.
%%%%%%%%%%%%%%%%%%%%%%%%%
\subsection{Thermodynamics and susceptibilities}\label{sec:npe_thermo}
A bulk viscosity calculation in neutrino-transparent $npe$ matter will require information about the thermodynamics of such as system, and so some relevant thermodynamic relations are reviewed here.  The strong and electromagnetic interactions are powerful enough to force the neutrons, protons, and electrons to move together as one fluid.  This matter produces neutrinos, but the neutrinos, participating only in the weak interaction, have a much longer MFP and this section will consider thermodynamic conditions where they escape the system.  The baryon number of the system $n_B$ is given by $n_B = n_n + n_p$ and and charge neutrality of the nuclear matter demands $n_e=n_p$, leaving just one independent particle species, chosen here to be the proton.  If the matter is in beta equilibrium, the proton fraction has a fixed value at a particular baryon density and temperature.

A beta equilibrium condition can be derived from any allowed reaction in the system, provided that all particles in the reaction are in thermal equilibrium.  Thus, in neutrino-trapped nuclear matter, the beta equilibrium condition is determined by examining the reaction $e^-+p\leftrightarrow n + \nu$, and through a standard thermodynamic procedure involving stoichiometric coefficients \cite{Schaffner-Bielich:2020psc}, the condition 
\begin{equation}
    \delta\mu_{\text{trapped}}\equiv \mu_n +\mu_{\nu} - \mu_p - \mu_e = 0 \label{eq:beq_nutrans}
\end{equation}
fixes the matter in chemical equilibrium.  To get the neutrino-transparent condition, it is common to set the neutrino chemical potential in Eq.~\ref{eq:beq_nutrans} to zero, since neutrino number is not conserved in neutrino-transparent matter, yielding the beta equilibrium condition
\begin{equation}
    \delta\mu \equiv \mu_n -\mu_p - \mu_e = 0.
\end{equation}
This equation will be sufficient for this chapter, however it does pick up corrections at finite temperature\footnote{Free-streaming neutrinos are not equivalent to a neutrino Fermi gas with zero chemical potential, so there is no reason to expect $\mu_n=\mu_p+\mu_e$ to be the proper beta equilibrium condition in neutrino-transparent matter.  The proper condition is found by equating the neutron-producing rates with the proton-producing rates \cite{Alford:2018lhf,Alford:2021ogv}.  However, it was recently found that using $\mu_n=\mu_p+\mu_e$ instead of the condition as finite temperature does not substantially change the bulk viscosity \cite{Alford:2023gxq}.\label{footnote:violate_beq_condition}}

The first law of thermodynamics in $npe$ matter is \cite{Schaffner-Bielich:2020psc,swendsen2020introduction}
\begin{equation}
    \mathop{dE} = -P\mathop{dV}+T\mathop{dS}+\mu_n\mathop{dN_n}+\mu_p\mathop{dN_p}+\mu_e\mathop{dN_e}.\label{eq:1st_law_extensive}
\end{equation}

Normalizing by the volume, and using the Euler equation
\begin{align}
    \varepsilon+P &= sT + \mu_nn_n+\mu_pn_p+\mu_en_e\nonumber\\
    &= sT + \mu_nn_B-n_p\delta\mu,
\end{align}
 (derived in, e.g.~\cite{swendsen2020introduction}) leads to
\begin{align}
    \mathop{d\varepsilon} &= T\mathop{ds}+\mu_n\mathop{dn_n}+\mu_p\mathop{dn_p}+\mu_e\mathop{dn_e}\nonumber\\
    &= T\mathop{ds}+\mu_n\mathop{dn_B}-\delta\mu\mathop{dn_p}.\label{eq:1st_law_intensive}
\end{align}
The Gibbs-Duhem equation, in some sense the compliment to the first law of thermodynamics, can be derived by taking the total derivative of the Euler equation and then replacing $\mathop{d\varepsilon}$ by Eq.~\ref{eq:1st_law_intensive}, leading to
\begin{align}
\mathop{dP}&=s\mathop{dT}+n_n\mathop{d\mu_n}+n_p\mathop{d\mu_p}+n_e\mathop{d\mu_e}\nonumber\\
&= s\mathop{dT}+n_B\mathop{d\mu_n}-n_p\mathop{d\delta\mu}.
\end{align}
In these derivations, I have made use of the conditions $\mathop{dn_B}=\mathop{dn_n}+\mathop{dn_p}$ and $\mathop{dn_e}=\mathop{dn_p}$.

Instead of normalizing the first law (Eq.~\ref{eq:1st_law_extensive}) by volume, it can be normalized by baryon number $N_B$, which is conserved (that is, $\mathop{dN_B}=0$).  This leads to the expression
\begin{equation}
    \mathop{d\left(\frac{\varepsilon}{n_B}\right)}=\frac{P}{n_B^2}\mathop{dn_B}+T\mathop{d\left(\frac{s}{n_B}\right)}-\delta\mu\mathop{dx_p}.
\end{equation}

Since it is more natural to specify temperature than entropy in the EoS RMF calculation, I will work at constant temperature in this chapter.  The above thermodynamic relations must be Legendre transformed \cite{swendsen2020introduction}, leading to the equations
\begin{align}
    \mathop{d\left(\varepsilon-sT\right)} &= \mu_n\mathop{dn_B}-s\mathop{dT}-\delta\mu\mathop{dn_p},\\
    \mathop{d\left(\frac{\varepsilon-sT}{n_B}\right)} &= \frac{P}{n_B^2}\mathop{dn_B}-\frac{s}{n_B}\mathop{dT}-\delta\mu\mathop{dx_p}.\label{eq:1st_law_per_baryon}
\end{align}
Many Maxwell relations can be derived from these relations, but the only important one for our purpose comes from Eq.~\ref{eq:1st_law_per_baryon}
\begin{equation}
    \frac{\partial P}{\partial x_p}\bigg\vert_{n_B,T}=-n_B^2\frac{\partial\delta\mu}{\partial n_B}\bigg\vert_{x_p,T}.
\end{equation}

In calculating the bulk viscosity resulting from small amplitude density oscillations, it is necessary to consider small perturbations around a state of chemical equilibrium.  The coefficients of the Taylor expansions will be written in terms of susceptibilities, which are properties of the nuclear EoS.  In degenerate nuclear matter, the susceptibilities are essentially unconstrained at present.  As a consequence (c.f.~Eq.~\ref{eq:zeta_max}), the maximum value of the subthermal bulk viscosity is also essentially unconstrained.

In this chapter, I will consider isothermal density oscillations, so the isothermal susceptibilities 
\begin{equation}
    A \equiv n_B\frac{\partial\delta\mu}{\partial n_B}\bigg\vert_{T,x_p} = -\frac{1}{n_B}\frac{\partial P}{\partial x_p}\bigg\vert_{n_B,T}\label{eq:A_susc_npe}
\end{equation}
and
\begin{equation}
    B \equiv \frac{1}{n_B}\frac{\partial \delta\mu}{\partial x_p}\bigg\vert_{n_B,T}\label{eq:B_susc_npe}
\end{equation}
are defined.  Expressions relating the isothermal and adiabatic (constant entropy per baryon) susceptibilities are provided in the appendix of \cite{Alford:2019qtm}.  Numerically, there is little difference between the two for densities above $n_0$ and temperatures below 10 MeV \cite{Alford:2019qtm}.

\begin{figure*}
  \centering
  \includegraphics[width=0.4\textwidth]{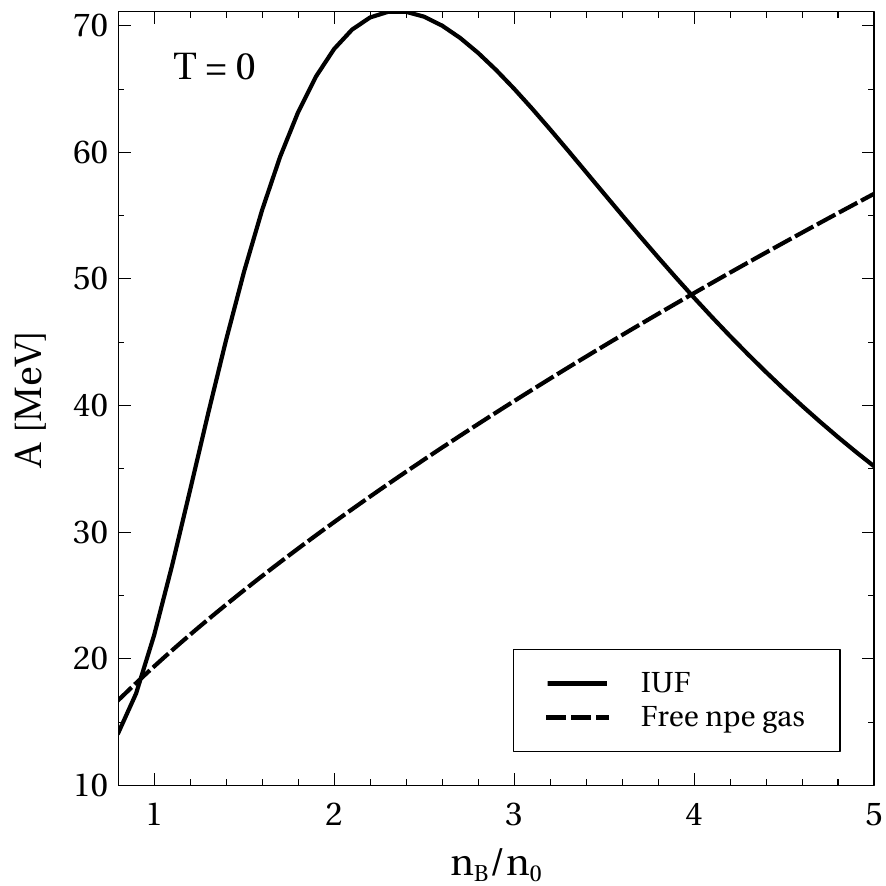}
  \includegraphics[width=0.4\textwidth]{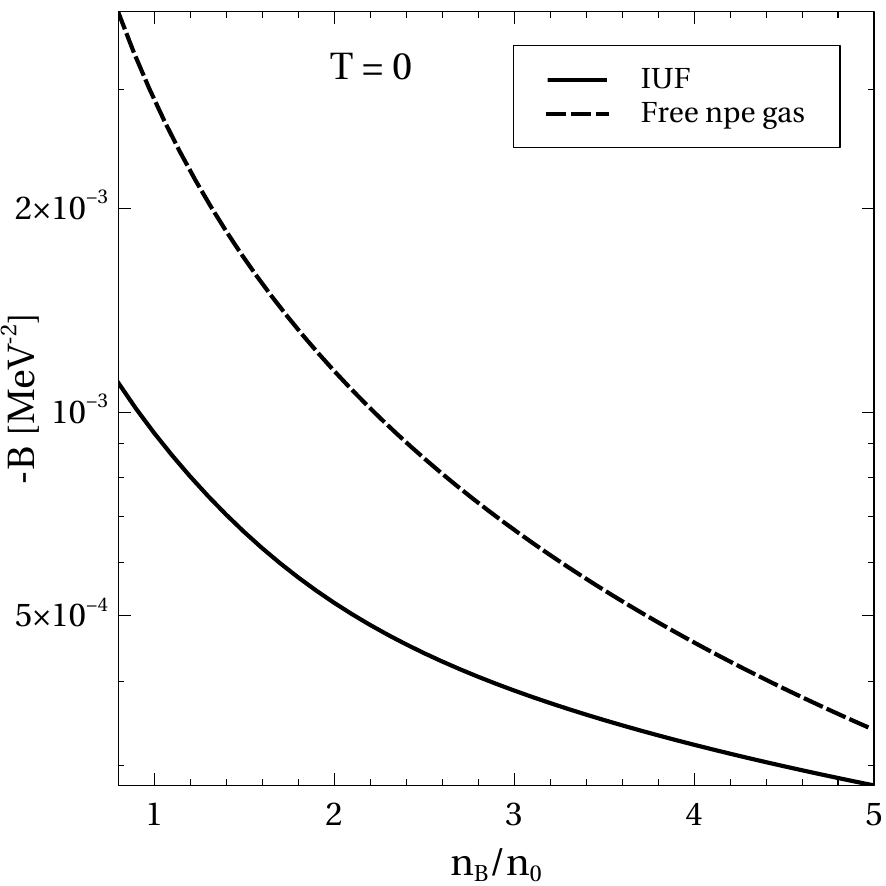}
  \caption{Isothermal susceptibilities in $npe$ matter, with the solid lines representing the IUF result and the dashed line representing the free Fermi gas result \cite{Alford:2010gw}.  The susceptibilities are calculated at zero temperature, but the temperature dependence is small.}
  \label{fig:A_B_susc}
\end{figure*}

In Fig.~\ref{fig:A_B_susc} are plotted the susceptibilities for the IUF EoS, compared with those calculated for a free Fermi gas (from table 2 in \cite{Alford:2010gw}).  The susceptibilities do not change significantly with temperature, at least in the neutrino-transparent regime (see, e.g., Fig.~3 in \cite{Alford:2019qtm}).
%%%%%%%%%%%%%%%%%%%%%%%%%%%%%%%%%%%%%%%%%%%%%
\subsection{Bulk viscosity and beta equilibration}\label{sec:npe_bv_and_beq}
\begin{figure}
  \centering
  \includegraphics[width=0.4\textwidth]{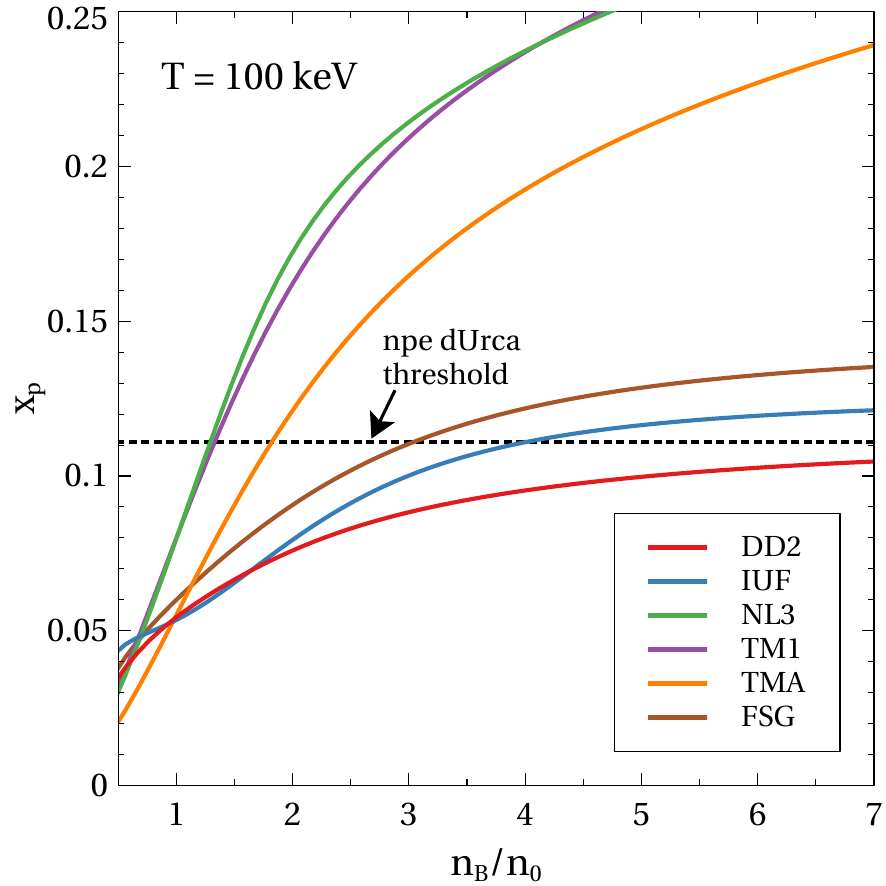}
  \caption{Beta-equilibrium proton fraction of several common $npe$ EoSs, as a function of density.  At zero temperature, the direct Urca process is kinematically allowed at densities where the proton fraction is greater than 1/9, depicted by the dotted black line.  Data obtained from CompOSE \cite{CompOSECoreTeam:2022ddl}.}
  \label{fig:xp}
\end{figure}
To calculate the bulk viscosity, consider a fluid element undergoing a small amplitude, sinusoidal change in its density
\begin{align}
    n_B(t) &= n_B + \Re{(\delta n_B e^{i\omega t})}\nonumber\\
    &= n_B + \delta n_B \cos{(\omega t)},\label{eq:nb_of_t}
\end{align}
where $\delta n_B \ll n_B$.  Because the beta equilibrium value of the proton fraction (plotted for several typical $npe$ EoSs in Fig.~\ref{fig:xp}) is a function of density, when the matter is compressed, the matter departs from beta equilibrium\footnote{If the beta-equilibrium value of the proton fraction does not change with density, the system is conformal and the bulk viscosity is zero.  Dense matter can be conformal at the particular density where the beta-equilibrium proton fraction is at a minimum or a maximum.  This situation is unlikely to occur in uniform $npe$ matter (Fig.~\ref{fig:xp} shows that $x_p$ monotonically increases with density), but can occur in neutrino-trapped matter \cite{Alford:2019kdw}.  Dense $npe$ matter that has zero bulk viscosity from Urca processes at a particular density would still have a (very small) collisional bulk viscosity \cite{Kolomeitsev:2014gfa} and thus is not \textit{truly} conformal.}.  In response, chemical reactions will try to bring the system to the new value of the beta-equilibrium proton fraction and thus the proton fraction will also be a periodic function
\begin{align}
    x_p(t) &= x_p^0 + \Re{(\delta x_p e^{i\omega t})}\nonumber\\
    &= x_p^0 + \Re{(\delta x_p)}\cos{(\omega t)} - \Im{(\delta x_p)}\sin{(\omega t)}.\label{eq:xp_of_t}
\end{align}
However, the chemical reactions only have a finite rate, and depending on how the reaction rate compares to the density oscillation frequency, there could be a phase lag between the proton fraction and baryon density.  The pressure of the nuclear matter is a function of both the density and the proton fraction (and the temperature, but this will be neglected) and will also undergo a harmonic oscillation 
\begin{align}
    P(t) &= P_0 + \Re{(\delta P e^{i\omega t})}\nonumber\\
    &= P_0+ \Re{(\delta P)}\cos{(\omega t)} - \Im{(\delta P)}\sin{(\omega t)}\label{eq:P_of_t}
\end{align}
with a potential phase lag with respect to the baryon density.  Bulk-viscous energy dissipation comes from a phase lag between the pressure and volume of a fluid element throughout an oscillation period.  Baryon density is an obvious proxy for volume ($n_B \equiv N_B/V)$ and the proton fraction is a proxy for the pressure.  The part of the pressure that depends on density will not cause dissipation, but the part of the pressure that depends on $x_p$ will give rise to a term with a phase lag with respect to the baryon density.  One should expect the bulk viscosity to be proportional to $\Im{(\delta P)}$, since that is the coefficient of the part of the pressure (Eq.~\ref{eq:P_of_t}) that is out of phase with the density.  Finally, when a sinusoidal density oscillation is imposed, the proton fraction, pressure, and beta equilibrium $\delta\mu$ all undergo harmonic oscillations, some potentially with a phase lag.  We should expect the temperature to act similarly, but, following Langer \& Cameron \cite{1969Ap&SS...5..213L}, temperature changes throughout an oscillation will be neglected throughout this chapter. 

The oscillation energy (per unit volume) lost due to compression or rarifaction in fluid motion (that is, motion with net divergence) is encapsulated in the bulk viscosity coefficient $\zeta$ via the relation
\begin{equation}
    \dfrac{\mathop{d\varepsilon_{\text{osc}}}}{\mathop{dt}} = - \zeta (\nabla \cdot \mathbf{v})^2.
\end{equation}
Using the continuity equation for a comoving fluid element
\begin{equation}
    \dfrac{\mathop{dn_B}}{\mathop{dt}} + n_B \nabla \cdot \mathbf{v}=0,
\end{equation}
and then averaging over one oscillation period, we find
\begin{equation}\label{eq:dedt1}
    \left\langle\dfrac{\mathop{d\varepsilon_{\text{osc}}}}{\mathop{dt}}\right\rangle = -\dfrac{1}{2}\left(\dfrac{\delta n_B}{n_B}\right)^2\omega^2\zeta.
\end{equation}
To finish deriving the bulk viscosity coefficient in terms of $\Im{(\delta P)}$, we calculate the energy dissipation in terms of the $PdV$ work done in one oscillation
\begin{align}
    \left\langle \dfrac{\mathop{d\varepsilon}_{\text{diss}}}{\mathop{dt}}\right\rangle &= \dfrac{\omega}{2\pi}\int_0^{2\pi/\omega}\mathop{dt}\dfrac{P(t)}{n_{B}(t)}\dfrac{\mathop{dn_{B}}(t)}{\mathop{dt}}\label{eq:dedt2}\\
    &=\dfrac{\omega^2}{2\pi}\left(\dfrac{\delta n_B}{n_B}\right)\Im{(\delta P)}\int_0^{2\pi/\omega}\mathop{dt}\sin^2{\left(\omega t\right)}\nonumber\\
    &= \dfrac{\omega}{2}\left(\dfrac{\delta n_B}{n_B}\right)\Im{(\delta P)}.\nonumber
\end{align}
Since $\left\langle\dot{\varepsilon}_{\text{osc}}\right\rangle=-\left\langle\dot{\varepsilon}_{\text{diss}}\right\rangle$ (dot denotes time derivative), the bulk viscosity is given by
\begin{equation}\label{eq:zeta_imP}
    \zeta = \left(\dfrac{n_B}{\delta n_B}\right)\dfrac{\Im{(\delta P)}}{\omega}.
\end{equation}
This expression is general, even for more complex situations where there are multiple equilibrating quantities.

The pressure $P(n_B,T,x_p)$ can be expanded around its equilibrium value $P_0$
\begin{align}
    P &= P_0 + \dfrac{\partial P}{\partial n_B}\bigg\vert_{T,x_p}\Re{(\delta n_B e^{i\omega t})}+\dfrac{\partial P}{\partial T}\bigg\vert_{n_B,x_p}\Re{(\delta T e^{i\omega t})}\nonumber\\
    &+\dfrac{\partial P}{\partial x_p}\bigg\vert_{T,n_B}\Re{(\delta x_p e^{i\omega t})},\nonumber\\
    &= P_0 + \frac{1}{\kappa_T}\frac{\delta n_B}{n_B}\cos{(\omega t)}-n_BA\Re{(\delta x_p)}\cos{(\omega t)}\nonumber\\
    &+n_BA\Im{(\delta x_p)}\sin{(\omega t)},\label{Eq:P_of_t_2}
\end{align}
where the definitions of the susceptibility (Eq.~\ref{eq:A_susc_npe}) and the isothermal compressibility \cite{CompOSECoreTeam:2022ddl}
\begin{equation}
    \kappa_T = \left( \left. n_B\dfrac{\partial P}{\partial n_B}\right\vert_{T,x_p}\right)^{-1}
\end{equation}
were invoked.  Hereafter, I drop the temperature dependence in expansions around beta equilibrium. 
 Matching sine and cosine terms between the two equations for the pressure (Eq.~\ref{eq:P_of_t} and \ref{Eq:P_of_t_2}) yields
\begin{subequations}
\begin{align}
    \Re{(\delta P)} &= \frac{1}{\kappa_T}\frac{\delta n_B}{n_B}-n_BA\Re{(\delta x_p)}\\
    \Im{(\delta P)} &= -n_B A \Im{(\delta x_p)}. \label{eq:imdP}
\end{align}
\end{subequations}
In order to calculate the bulk viscosity, the imaginary part of the proton fraction oscillation must be found.

The proton fraction evolves through flavor-changing interactions.  In neutrino transparent $npe$ matter, the direct Urca (dUrca) processes
\begin{subequations}
\begin{align}
    n &\rightarrow p + e^- + \bar{\nu}_e \quad \text{dUrca neutron decay}\\
    e^- + p &\rightarrow n + \nu_e \quad \text{dUrca electron capture}
\end{align}
\end{subequations}
and the modified Urca (mUrca) processes
\begin{subequations}
\begin{align}
    N+n &\rightarrow N+p + e^- + \bar{\nu}_e \quad \text{mUrca neutron decay}\\
    N+e^- + p &\rightarrow N+n + \nu_e \quad \text{mUrca electron capture}
\end{align}
\end{subequations}
act to establish beta equilibrium.  Because the matter is assumed to be neutrino transparent (due to the long neutrino MFP at low temperatures), neutrinos and antineutrinos are not allowed to be on the left-hand side of any reaction.  

At temperatures below about 1 MeV, the $npe$ matter is so strongly degenerate that the Urca rates above can be calculated in the Fermi surface (FS) approximation, where only particles within energy $T$ of the Fermi energy are allowed to participate in the reaction.  This approximation illuminates the strong temperature dependence of rates in degenerate nuclear matter.  In this approximation, the direct Urca process is only kinematically allowed for densities such that $p_{Fn} < p_{Fp}+p_{Fe}$, which corresponds to proton fractions larger than $1/9$ in $npe$ matter \cite{Schmitt:2010pn}.  For the IUF EoS, the direct Urca threshold is near $n_B = 4n_0$ (c.f.~Fig.~\ref{fig:xp}), and thus below $4n_0$, only the modified Urca process is there to help establish beta equilibrium.  As the temperature rises above about 1 MeV, the Boltzmann suppression of reaction participants that lie far from the Fermi surface decreases, causing the direct Urca threshold density to blur (c.f.~Fig.~1 in \cite{Alford:2019qtm}) as the reaction phase space opens up.  This effect is important for temperatures above 1 MeV \cite{Alford:2018lhf}, but I will neglect it in this chapter so that I can focus on other aspects of the bulk viscosity calculation.

The proton fraction evolves according to the equation
\begin{align}
    n_B\frac{\mathop{dx_p}}{\mathop{dt}} &= \Gamma_{nn\rightarrow npe\nu}-\Gamma_{nep\rightarrow nn\nu}\nonumber\\
    &=\frac{1}{5760\pi^9}G^2g^2_Af^4\frac{m_n^4}{m^4_{\pi}}\frac{p^4_{Fn}p_{Fp}}{(p^2_{Fn}+m^2_{\pi})^2}\theta_n\delta\mu\label{eq:suprathermal_murca}\\ 
    &\times (1835\pi^6T^6+945\pi^4\delta\mu^2T^4+105\pi^2\delta\mu^4T^2+3\delta\mu^6).\nonumber\\
    &\approx \lambda_{npe} \delta\mu \quad \text{(subthermal limit)},\label{eq:xp_evolution}
\end{align}
with
\begin{align}
    \theta_n=\begin{cases}
    1 & p_{Fn}>p_{Fp}+p_{Fe}\\
    1-\dfrac{3}{8}\dfrac{(p_{Fp}+p_{Fe}-p_{Fn})^2}{p_{Fp}p_{Fe}} & p_{Fn}<p_{Fp}+p_{Fe}\,
    \end{cases}\label{eq:theta_n}
\end{align}
and\footnote{In reality, this is an older version of the modified Urca rate (used in, e.g.~Alford \& Harris \cite{Alford:2019qtm}).  The factor of $m_n^4$ should be replaced by $(E^*_{Fn})^3E^*_{Fp}$ where $E^*_{Fi}=\sqrt{p_{Fi}^2+m_*^2}$ \cite{Alford:2021ogv}.  I use the older expression here because while it overestimates the modified Urca rate, it gets the bulk-viscous resonance to be at the same temperature as more realistic calculations \cite{Alford:2023gxq}.}
\begin{equation}
    \lambda_{npe} = \frac{367}{1152\pi^3}G^2g^2_Af^4\frac{m_n^4}{m^4_{\pi}}\frac{p^4_{Fn}p_{Fp}}{(p^2_{Fn}+m^2_{\pi})^2}T^6\theta_n.\label{eq:lambda_npe}
\end{equation}
In this chapter, all $\lambda$ coefficients are positive.  In the above formulas, $G^2 \equiv G_F^2\cos^2{\theta_c} = 1.29\times10^{-22} \text{ MeV}^{-4}$, where $G_F$ is the Fermi coupling constant and $\theta_C$ is the Cabibbo angle.  The axial vector coupling constant $g_A = 1.26$ and $f\approx 1$.  
Now, $\delta\mu$ can be expanded around chemical equilibrium 
\begin{align}
    \delta\mu &=  \dfrac{\partial \delta\mu}{\partial n_B}\bigg\vert_{T,x_p}\Re{(\delta n_B e^{i\omega t})}+  \dfrac{\partial \delta\mu}{\partial x_p}\bigg\vert_{n_B,T}\Re{(\delta x_p e^{i\omega t})}\nonumber\\
    &\approx A\frac{\delta n_B}{n_B}\cos{(\omega t)}+n_BB\big[\Re{(\delta x_p)}\cos{(\omega t)}\nonumber\\
    &-\Im{(\delta x_p)}\sin{(\omega t)}\big].\label{eq:deltamu_of_t}
\end{align}
Plugging Eqs.~\ref{eq:xp_of_t} and \ref{eq:deltamu_of_t} into Eq.~\ref{eq:xp_evolution}, and then matching the coefficients of the sine and cosine terms, a system of two equations for $\Re{(\delta x_p)}$ and $\Im{(\delta x_p)}$ is obtained, the solution of which is
\begin{subequations}
\begin{align}
    \Re{(\delta x_p)} &= -\left(\frac{\delta n_B}{n_B}\right)\frac{AB\lambda^2}{n_B\left(\omega^2+B^2\lambda^2\right)}\\
    \Im{(\delta x_p)} &= -\left(\frac{\delta n_B}{n_B}\right)\frac{A\lambda\omega}{n_B\left(\omega^2+B^2\lambda^2\right)}\label{eq:imdxp}
\end{align}
\end{subequations}
Combining Eqs.~\ref{eq:zeta_imP}, \ref{eq:imdP}, and \ref{eq:imdxp}, we obtain the subthermal bulk viscosity in $npe$ matter
\begin{equation}
    \zeta_{npe} = \frac{A^2\lambda}{\omega^2+B^2\lambda^2} = \frac{A^2}{\vert B\vert}\frac{\gamma}{\omega^2+\gamma^2},\label{eq:bv_npe}
\end{equation}
where the equilibration rate $\gamma$ is defined as
\begin{equation}
    \gamma \equiv \vert B\vert\lambda. \label{eq:gamma_npe}
\end{equation}
The expressions for the real and imaginary parts of $\delta x_p(t)$ and $\delta P(t)$, solved for in pursuit of the bulk viscosity, can be used to obtain\footnote{$B$ is assumed to be negative in this calculation, as it is (right panel of Fig.~\ref{fig:A_B_susc}).} $x_p(t)$, $P(t)$, and $\delta\mu(t)$
\begin{widetext}
\begin{subequations}
\begin{align}
    x_p(t) &= x_p^0 - \left(\frac{\delta n_B}{n_B}\right)\frac{A}{Bn_B}\frac{\gamma}{\omega^2+\gamma^2}\left[\gamma\cos{(\omega t)}+\omega\sin{(\omega t)}\right],\label{eq:xp_of_t_specific}\\
    P(t) &= P_0 + \left(\frac{\delta n_B}{n_B}\right)\left\{ \frac{1}{\kappa_T}\cos{(\omega t)}+\frac{A^2}{B}\frac{\gamma}{\omega^2+\gamma^2}\left[ 
\gamma\cos{(\omega t)}+\omega\sin{(\omega t)}  \right]  \right\},\label{eq:P_of_t_function}\\
    \delta\mu(t) &= A\left(\frac{\delta n_B}{n_B}\right)\frac{\omega}{\omega^2+\gamma^2}\left[\omega\cos{(\omega t)}-\gamma\sin{(\omega t)}\right].\label{eq:dmu_of_t}
\end{align}
\end{subequations}
\end{widetext}
\begin{figure*}
  \centering
  \includegraphics[width=.4\textwidth]{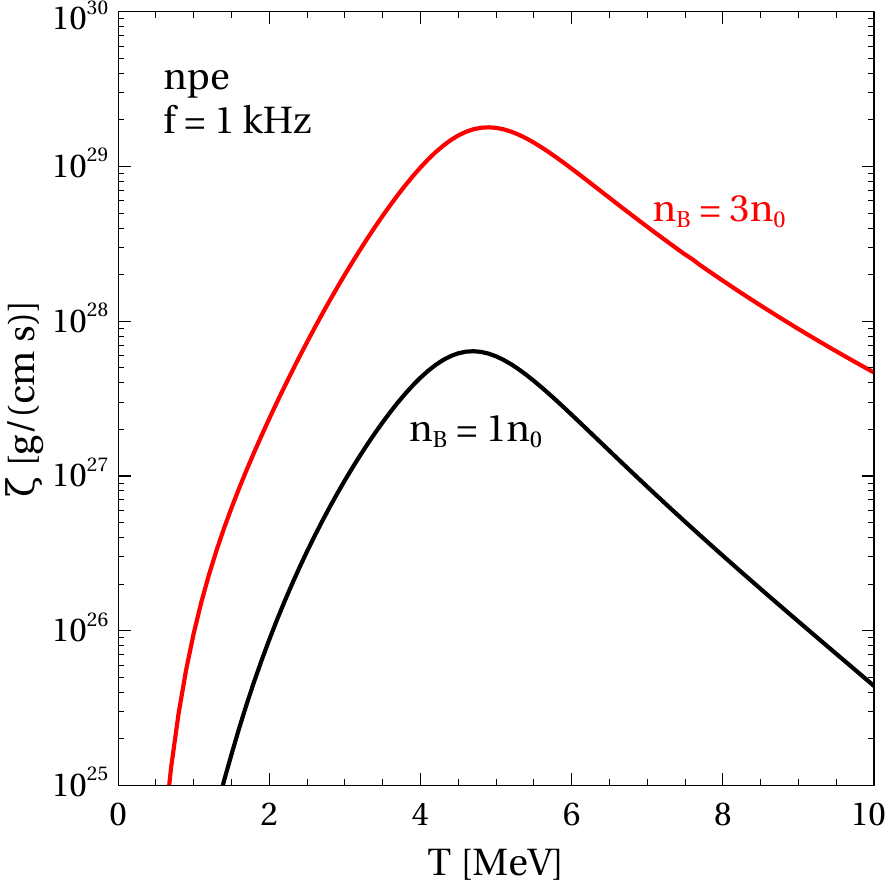}
  \includegraphics[width=.4\textwidth]{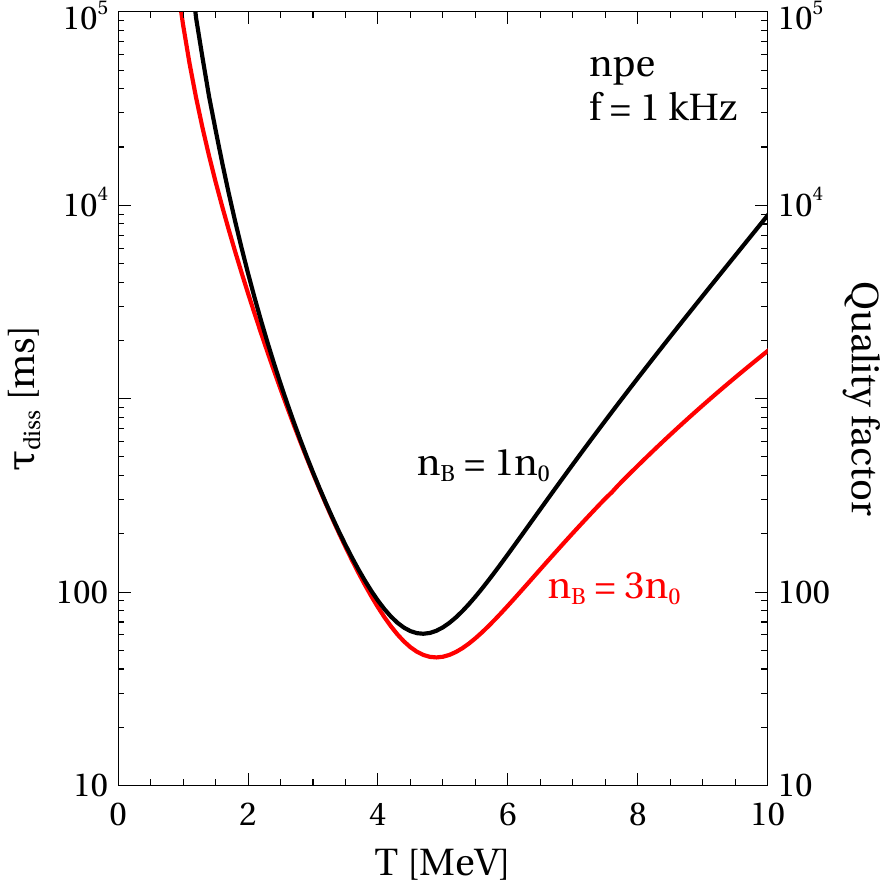}
  \caption{Left: Bulk viscosity of neutrino-transparent $npe$ matter undergoing a density oscillation with angular frequency $\omega = 2\pi\times 1\text{ kHz}$.  Right: Timescale over which bulk viscosity dissipates kinetic energy from the oscillation (left axis), and the quality factor of the oscillation (right axis).}
  \label{fig:npe_bv}
\end{figure*}

The bulk viscosity of $npe$ matter is plotted as a function of temperature in the left panel of Fig.~\ref{fig:npe_bv}.  The density oscillation frequency is chosen to be 1 kHz ($\omega = 2\pi\times 1\text{ kHz}$), typical of neutron stars and neutron star mergers\footnote{The frequency of a sound wave in a neutron star can be estimated by dividing the distance scale (10 km) by the speed of sound in dense matter (0.1) \cite{Tews:2018kmu}, which yields a timescale of 0.33 ms, or a linear frequency of 3 kHz.\label{footnote:frequencyestimate}  Sec.~\ref{sec:postmerger} shows evidence for this oscillation frequency from numerical simulations.}.  The key feature of the bulk viscosity as a function of temperature is the presence of one resonant maximum, occurring at the temperature where the beta equilibration rate $\gamma$ matches the density oscillation frequency $\omega$.  The beta equilibration rate in $npe$ matter is monotonic with temperature ($\lambda_{npe}\sim T^6$, Eq.~\ref{eq:lambda_npe}), and there are no additional parallel beta equilibration channels (as there will be in $npe\mu$ matter) and thus there is only one resonance.  

At low temperatures, the beta equilibration rate is very slow compared to the density oscillation frequency, and thus the matter barely equilibrates upon compression, and therefore the system changes very little throughout the oscillation (beyond the non-dissipative increase and decrease of the pressure due to the density change alone).  In this regime, the bulk viscosity is approximately
\begin{equation}
    \zeta_{npe}^{\gamma\ll\omega} \approx \frac{A^2}{\vert B\vert}\frac{\gamma}{\omega^2},\label{eq:zeta_cold}
\end{equation}
which is small because $\gamma\ll\omega$.  This is the form of the bulk viscosity used in studying cold neutron stars, as described in Sec.~\ref{sec:history} (e.g.~\cite{Haensel:2000vz,Haensel:2001mw,Haensel:2001em,Lindblom:2001hd,Yakovlev:2018jia}).  The bulk viscosity at fixed frequency $\omega$ is proportional to the equilibration rate\footnote{Because the bulk viscosity in this limit is proportional to the beta equilibration rate, an early goal in bulk viscosity research was to find ``the fastest'' equilibration process, as it would lead to the largest bulk viscosity.  However, it was known that some processes were ``too fast'', that is, faster than the millisecond oscillation timescales and thus on the other side of their resonant peak (e.g.~$n+n \leftrightarrow p + \Delta^-$ \cite{1969Ap&SS...5..213L}).}.

At high temperatures, the Urca rate is very fast, and when the density changes, the matter essentially never departs from chemical equilibrium and there is little dissipation.  In the high-temperature limit, the bulk viscosity is approximately
\begin{equation}
\zeta_{npe}^{\gamma\gg\omega} \approx \frac{A^2}{\vert B\vert}\frac{1}{\gamma}.    
\end{equation}
It is independent of the oscillation frequency, and as the rate $\gamma$ grows faster with increasing temperature, the bulk viscosity decreases.  The high-temperature limit is, of course, the low-frequency limit, and this expression for the bulk viscosity describes its behavior with respect to arbitrarily slow density oscillations.  It is also common to refer to this ``zero-frequency'' bulk viscosity coefficient as \textit{the} bulk viscosity, from which a frequency-dependent ``effective bulk viscosity'' $\zeta_{\text{eff}}(\omega)$ (that is, what I call \textit{the} bulk viscosity) can be obtained \cite{Gavassino:2023eoz}.

Assuming that the susceptibilities do not depend on temperature, the maximum bulk viscosity occurs at the temperature where $\gamma = \omega$ and is 
\begin{equation}
    \zeta_{npe,\text{ max}} = \frac{1}{2\omega}\frac{A^2}{\vert B\vert}.\label{eq:zeta_max}
\end{equation}
Thus, in the simple case where there is only one chemical equilibration channel, the location of the maximum bulk viscosity is controlled by the Urca rate, and the strength of the bulk viscosity at resonance is controlled by the susceptibilities, which are properties of the EoS.  As we see from the left panel of Fig.~\ref{fig:npe_bv}, the bulk viscosity increases with density, which is due to the fact that $A$ increases from $n_0$ to $3n_0$ as $\vert B\vert$ decreases in the same interval (see Fig.~\ref{fig:A_B_susc}).

\begin{figure*}
  \centering
  \includegraphics[width=0.4\textwidth]{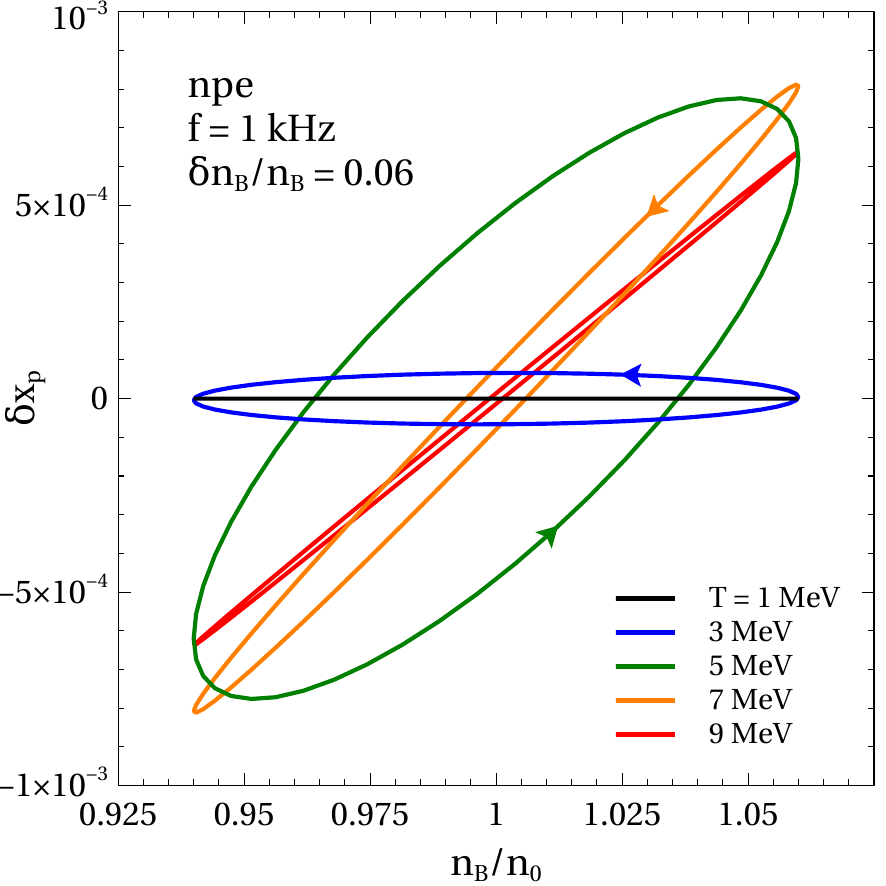}
  \includegraphics[width=0.4\textwidth]{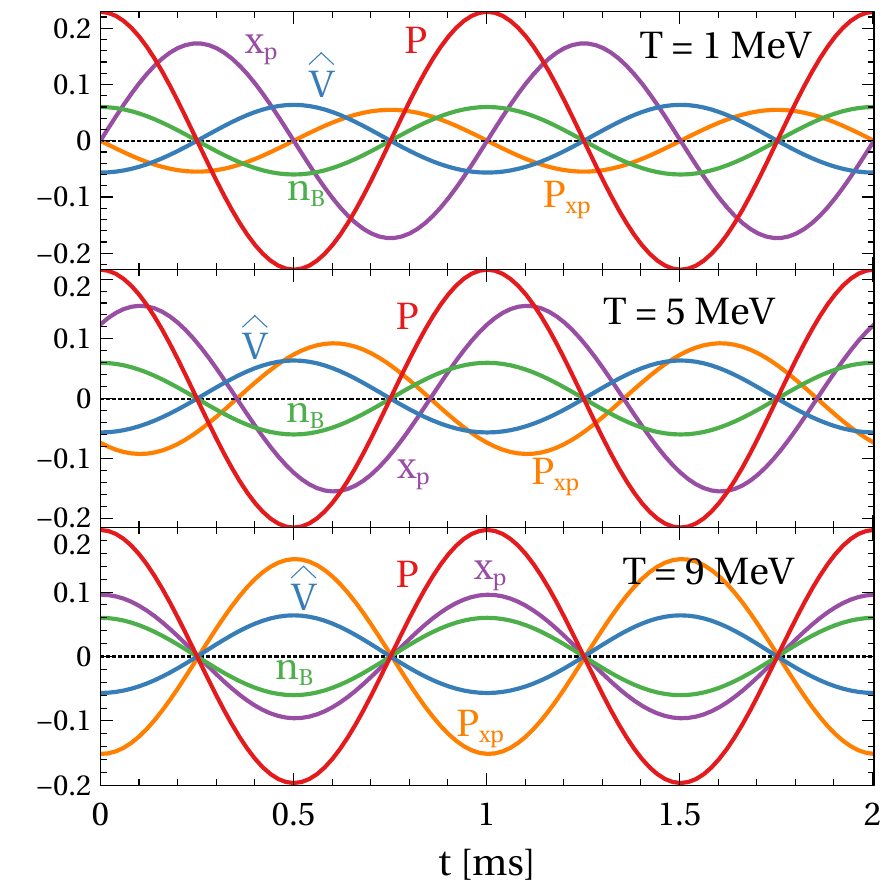}
  \caption{Left: Route through the $x_pn_B$ plane of a fluid element in one oscillation period.  Right: $P(t)$, $n_{B}(t)$, $\hat{V}(t)\equiv 1/n_{B}(t)$, $x_p(t)$, and $P_{xp}$ throughout two cycles, for three different temperatures.  The normalization in the right panel is arbitrary - the focus is the phase shifts between the various curves.}
  \label{fig:xpvsnB_and_xpnBPVvst}
\end{figure*}
\begin{figure*}
  \centering
  \includegraphics[width=0.4\textwidth]{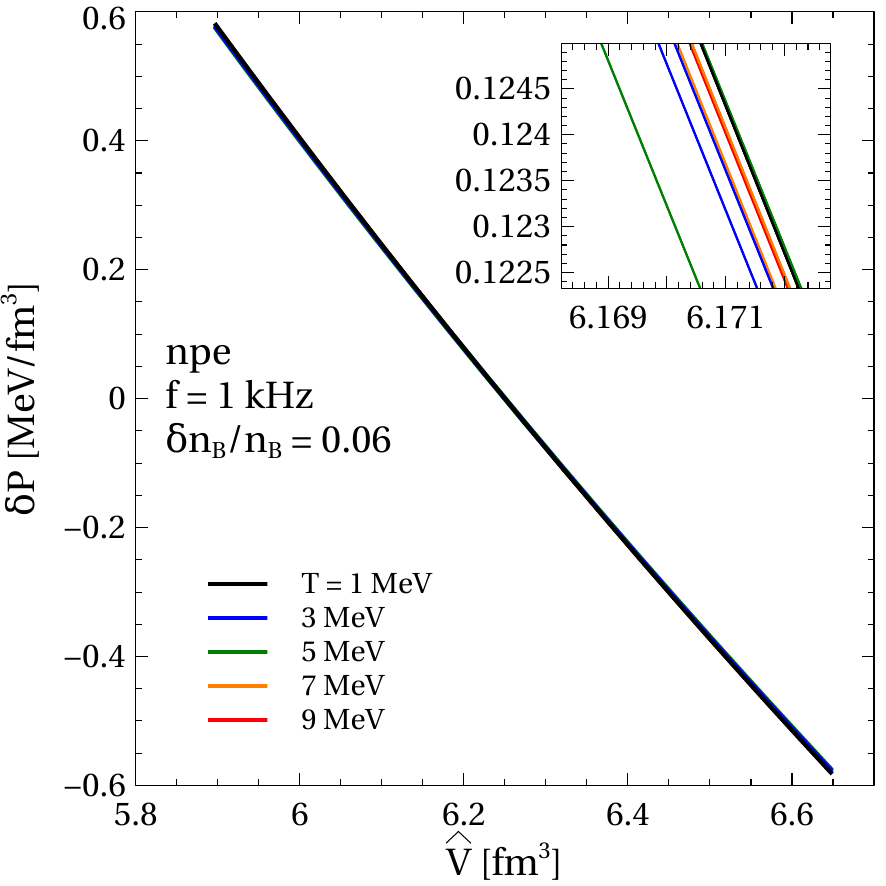}
  \includegraphics[width=0.4\textwidth]{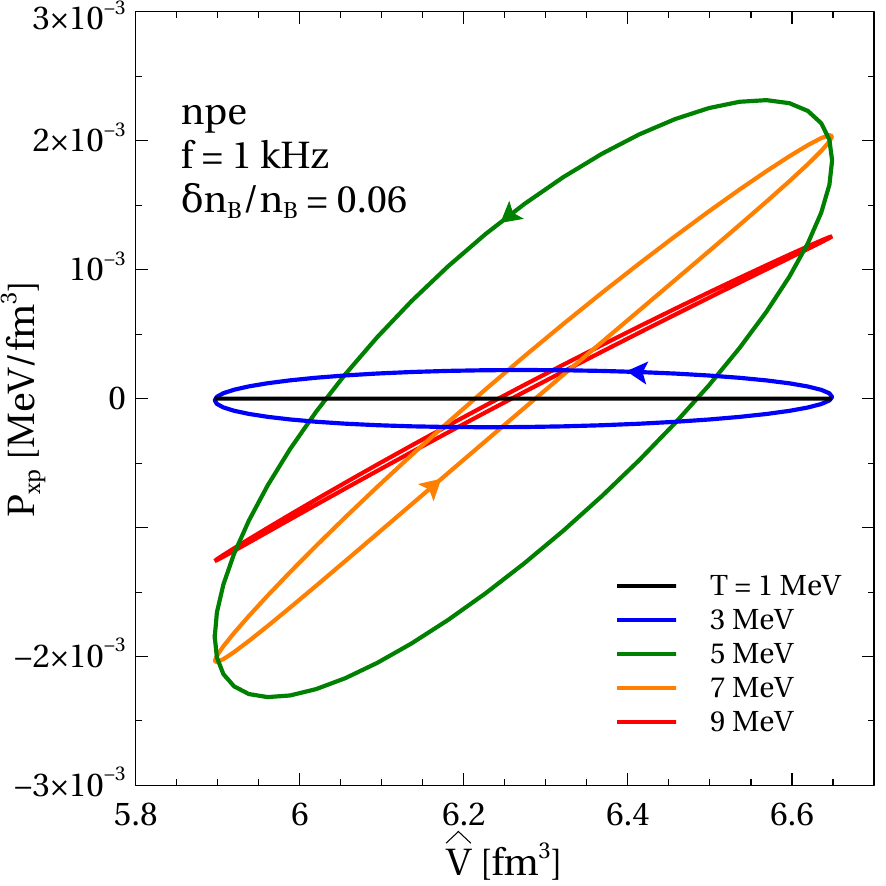}
  \caption{Left: Path through the $PV$ plane of a fluid element during one cycle of the density oscillation.  All five curves are overlapping due to the fact that the pressure depends much more on the density than the proton fraction, so the inset panel zooms in on a section of the curve to show the five individual curves.  Right: Path through the $P_{xp}V$ plane of a fluid element during one cycle of the density oscillation, therefore showing only the part of the pressure that depends on $x_p$, removing the part of the pressure that depends on $n_B$ as it does not contribute to the $PdV$ work.}
  \label{fig:PV_curves}
\end{figure*}

The resonance behavior of bulk viscosity can be further explored by looking at the path of a fluid element in the $x_pn_B$ (left panel of Fig.~\ref{fig:xpvsnB_and_xpnBPVvst}) or $PV$ (Fig.~\ref{fig:PV_curves}) plane, an analysis first put forward by Langer \& Cameron \cite{1969Ap&SS...5..213L}, but also discussed by Muto \textit{et al.}~\cite{1993PThPS.112..221M} and Camelio \textit{et al.}~\cite{Camelio:2022ljs}.  First, note that the pressure $P = P(n_B,T,x_p)$ is a function of the three thermodynamic variables, (the $T$-dependence is neglected, as discussed previously).  The density dependence of the pressure is very strong, and while the $x_p$ dependence is relatively weak, it is the source of the bulk viscosity effect.  It is natural to separate the pressure Eq.~\ref{eq:P_of_t} throughout the density oscillation into two pieces
\begin{equation}
P(t) = P_0 + \left(\frac{\delta n_B}{n_B}\right)\frac{1}{\kappa_T}\cos{(\omega t)}+P_{xp},    
\end{equation}
which, comparing with Eq.~\ref{eq:P_of_t}, serves as a definition of $P_{xp}$, the part of the pressure that varies due to chemical equilibration.

At one extreme, infinitely slow reactions, 
\begin{equation}
    P_{\gamma=0}(t)=P_0 + (\delta n_B/n_B)\kappa_T^{-1}\cos{(\omega t)}.
\end{equation}
At the other extreme, with infinitely fast equilibration,
\begin{equation}
    P_{\gamma\rightarrow\infty}(t) = P_0 + (\delta n_B/n_B)(\kappa_T^{-1}+A^2/B)\cos{(\omega t)}
\end{equation}
In both cases, the pressure is in phase with the density oscillation.  Examining these two limits, an interpretation for $A^2/B$, the prefactor of the bulk viscosity, becomes clear.  The inverse compressibility $\kappa_T^{-1}$ is the pressure increase caused by a density increase, with no chemical equilibration.  With infinitely fast equilibration, the inverse compressibility is corrected by an amount $A^2/B$.  Expanding out the expression, the incompressibility in the $\gamma\rightarrow\infty$ limit is
\begin{align}
\kappa_T^{-1}+\frac{A^2}{B} &= n_B\bigg( \dfrac{\partial P}{\partial n_B}\bigg\vert_{T,x_p}\\
&-\dfrac{\partial P}{\partial x_p}\bigg\vert_{T,n_B}  \dfrac{\partial x_p}{\partial \delta\mu}\bigg\vert_{T,n_B}  \dfrac{\partial \delta\mu}{\partial n_B}\bigg\vert_{T,x_p}   \bigg).\nonumber
\end{align}
This means that the prefactor of the bulk viscosity, the factor that determines the value it attains at resonance, is related to the difference between the incompressibility of the matter with infinitely fast chemical equilibration and the incompressibility with infinitely slow chemical equilibration.  This result makes clear the connection between the formalism described here and that of, for example, Lindblom \& Owen \cite{Lindblom:2001hd}, where they find the bulk viscosity is proportional to the difference of the adiabatic indices with infinitely fast or slow equilibration (itself a relativistic generalization of the Landau \& Lifshitz \cite{1987flme.book.....L} and Jones \cite{1970ApL.....5...33J} results involving the difference of sound speeds $c_{\infty}^2-c_0^2$).

The area enclosed by the path of a fluid element in the $PV$ plane (left panel of Fig.~\ref{fig:PV_curves}) indicates the work done in one cycle, and the direction the path is traversed indicates if the work is done on the system by the environment, or on the environment by the system \cite{1969Ap&SS...5..213L,1956ther.book.....F}.  Because dense matter is highly incompressible, the pressure depends strongly on the density and thus the $PV$ curve is close to a line.  A better representation of the $PdV$ work is seen by plotting the path of a fluid element in the $P_{xp}V$ plane (right panel of Fig.~\ref{fig:PV_curves}), which subtracts off the part of the pressure that does not contribute to the $PdV$ work done - that is, the area of a curve in the $PV$ plane is the same as the area in the $P_{xp}V$ plane.  The $x_pn_B$ plane (left panel of Fig.~\ref{fig:xpvsnB_and_xpnBPVvst}) gives similar information, as $x_p$ is essentially a proxy of $P_{xp}$, as $n_B$ is of $V$.
 
First consider a fluid element at $T=1\text{ MeV}$.  As the density is increased, the matter is pushed out of beta equilibrium, and wants to increase the proton fraction to reestablish the equilibrium.  However, the Urca processes are much too slow to do so within the duration of the oscillation, and thus the proton fraction remains very close to its original value.  In the density-proton fraction plane, the trajectory of one oscillation is very close to a straight line - the density increases and then decreases again, retracing its path.  The path of the fluid element through the $P_{xp}V$ plane is similar, enclosing very little area and thus dissipating little energy.

A fluid element with temperature 5 MeV undergoing a 1 kHz density oscillation is very close to experiencing the resonant behavior of bulk viscosity.  As the fluid element is compressed, the Urca process slowly turns on and adjusts the proton fraction, though lagging behind the density oscillation.  The two are nearly completely out of phase.  This leads the fluid element to traverse an ellipse with large area in the $x_pn_B$ plane, as well as the $P_{xp}V$ plane, and much energy is dissipated.  It is clear that the curve in the $PV$ plane is traversed counterclockwise, and therefore the environment does work on the system.  That is, the mechanism that causes the density oscillation does work on the fluid element, causing the oscillation to lose energy and the fluid element to heat up (ignoring the resulting enhancement of the neutrino emission which, in the subthermal regime, results in net cooling).

At higher temperatures, like $T=9\text{ MeV}$, upon compression, the fluid element almost instantaneously adjusts the proton fraction to reestablish beta equilibrium at the new density.  Pressure, proton fraction, and density are all in phase with each other and the energy dissipation is very small.

Plotted in the right panel of Fig.~\ref{fig:npe_bv} is the dissipation timescale of a density oscillation due to bulk viscosity.  The energy density of a density oscillation in nuclear matter is \cite{Alford:2019qtm}
\begin{equation}
    \varepsilon = \frac{1}{2}\left(\delta n_B\right)^2\frac{\partial^2\varepsilon}{\partial n_B^2}\bigg\vert_{x_p,T} = \frac{1}{2\kappa_T}\left(\frac{\delta n_B}{n_B}\right)^2.
\end{equation}
The rate of energy dissipation is directly related to the bulk viscosity, through Eq.~\ref{eq:dedt1}.  Thus a dissipation timescale can be obtained 
\begin{equation}
    \tau_{\text{diss}} \equiv \frac{\varepsilon}{\mathop{d\varepsilon}/\mathop{dt}}=\frac{1}{\kappa_T\omega^2\zeta}.
\end{equation}
The quality factor of the oscillation, or the number of oscillations before the oscillation energy drops by an e-fold \cite{taylor2005classical}, is \cite{Lindblom:2001hd}
\begin{equation}
    Q \equiv \frac{\omega}{2\pi}\tau_{\text{diss}}.
\end{equation}
Fig.~\ref{fig:npe_bv} indicates that density oscillations in $npe$ matter can be damped by bulk viscosity in as little as 30 ms.  Such a 1 kHz oscillation would have a quality factor of 30.  In more extensive scans of the thermodynamic conditions, Alford \& Harris \cite{Alford:2019qtm} and Alford, Haber, \& Zhang \cite{Alford:2023gxq} found dissipation times as little as 5 ms.  

\subsection{Validity of subthermal approximation}
In truncating the expression for the beta equilibration rate $\Gamma_{nn\rightarrow npe\nu}-\Gamma_{nep\rightarrow nn\nu}$ to first order in the departure from beta equilibrium, $\delta\mu$, we gain the ability to solve for the bulk viscosity exactly.  But, when using the subthermal bulk viscosity formalism to consider a density oscillation of a specific amplitude $\delta n_B$, one must be careful that $\delta\mu$ remains small compared to $T$.

\begin{figure}
  \centering
  \includegraphics[width=0.5\textwidth]{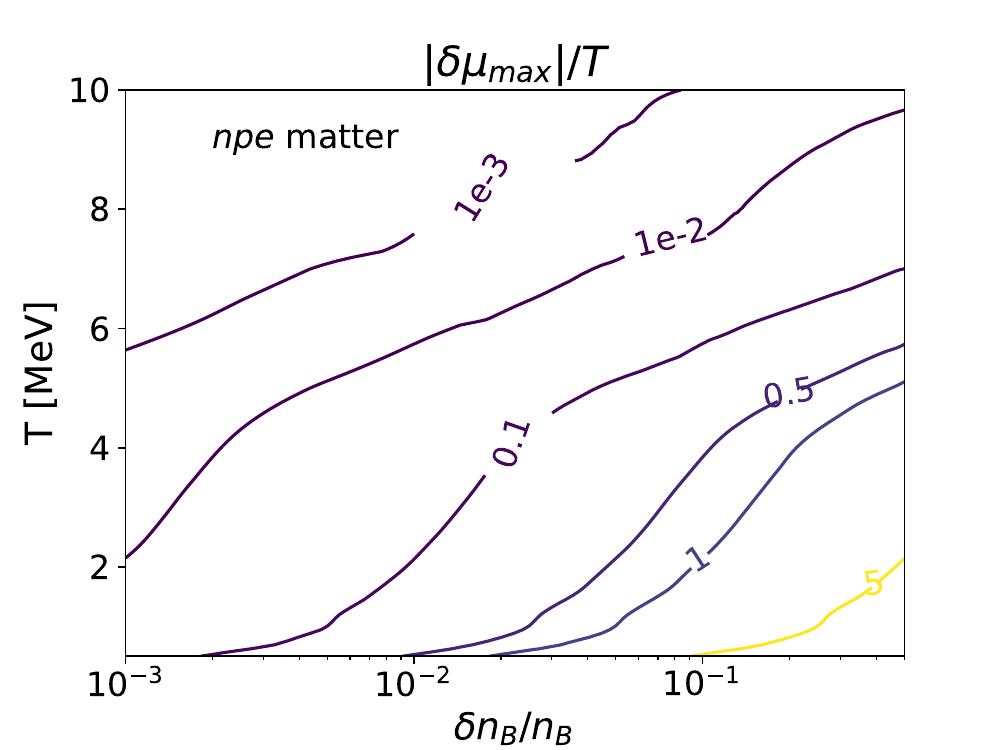}
  \caption{Contour plot of $\delta\mu(t)/T$ as a function of temperature and density oscillation amplitude.  If $\vert\delta\mu\vert/T\ll 1$, the oscillation is subthermal and the formalism developed in this chapter is applicable.  For larger density oscillations where $\delta\mu$ nears or exceeds $T$, the curve given in this figure and Eq.~\ref{eq:xp_evolution} no longer accurately describe the oscillation, and one should instead use the full expression Eq.~\ref{eq:suprathermal_murca}.}
  \label{fig:dmu_over_t}
\end{figure}

The maximum value of $\delta\mu$ in an oscillation cycle can be found from Eq.~\ref{eq:dmu_of_t}, and is given by
\begin{equation}
    \delta\mu_{\text{max}} = A\left(\frac{\delta n_B}{n_B}\right)\frac{\omega}{\sqrt{\omega^2+\gamma^2}.}\label{eq:dmu_max_npe}
\end{equation}
In a cold system, where $\gamma\ll\omega$, this reduces to $\delta\mu_{\text{max}}=A(\delta n_B/n_B)$ and thus it is ``easy'' to deviate strongly from the subthermal regime if $\delta n_B/n_B$ is too large.  In a hot system, where $\gamma\gg\omega$, $\delta\mu_{\text{max}}=A(\delta n_B/n_B)(\omega/\gamma)$, which remains small because the system is efficiently beta equilibrated and thus $\delta\mu$ is unable to depart significantly from zero.  The ratio $\delta\mu_{\text{max}}/T$ is plotted in the $T(\delta n_B/n_B)$ plane in Fig.~\ref{fig:dmu_over_t}.  All points $\{\delta n_B/n_B,T\}$ where $\delta\mu_{\text{max}}/T\ll 1$ lie in the subthermal regime. 

In matter with a temperature of 1 MeV, one leaves the subthermal regime if the density oscillation amplitude exceeds a few percent of the background density!  Arras \& Weinberg observed the suprathermal nature of the bulk viscosity in cold neutron stars in their study of beta reactions in the neutron star inspiral \cite{Arras:2018fxj}.  Alternatively, at a higher temperature like 8 MeV, the subthermal approximation is valid for $\delta n_B$ of the same order as $n_B$.  The bulk viscosity calculation as formulated here requires $\delta n_B\ll n_B$, or else the notion of expanding around an equilibrium state falls apart.  If the density oscillations are no longer small, which does seem to be the case in certain situations in neutron star mergers, then the reaction rates must be directly implemented in the simulation and evolved with the hydrodynamics, instead of ignored and encapsulated in a pre-calculated bulk-viscous contribution to the pressure.  
%%%%%%%%%%%%%%%%%%%%%%%%%%%%%%%%%%%%%%%%%%%%%%%%%%%%%%%%%
\section{Bulk viscosity in dense matter containing muons}\label{sec:npemu_bv}
\subsection{Thermodynamics and susceptibilities}
Adding muons to the standard $npe$ matter modifies properties of the EoS like the pressure and particle fractions, but also introduces new chemical equilibration channels.  In $npe\mu$ matter, the baryon density is still given by $n_B = n_n + n_p$ and the charge neutrality condition is now $n_p=n_e+n_{\mu}$.  Therefore $npe\mu$ matter has two independent particle species, chosen here to be the proton and muon.  There are now two beta equilibrium conditions (with all of the same issues observed in footnote \ref{footnote:violate_beq_condition}, as all flavor-changing reactions involve neutrinos or antineutrinos, which are free-streaming),
\begin{subequations}
\begin{align}
    \delta\mu_1 &\equiv \mu_n-\mu_p-\mu_e=0\label{eq:npemu_beq_1}\\
    \delta\mu_2 &\equiv \mu_n-\mu_p-\mu_{\mu}=0.\label{eq:npemu_beq_2}
\end{align}
\end{subequations}
Another natural choice would be $\mu_e=\mu_{\mu}$, but this condition is already implied by the two chosen conditions.  The thermodynamic formulas from Sec.~\ref{sec:npe_thermo} pick up terms with muons.  I will present them quickly here.  The first law of thermodynamics is
\begin{align}
    \mathop{dE} &= -P\mathop{dV}+T\mathop{dS}\nonumber\\
&+\mu_n\mathop{dN_n}+\mu_p\mathop{dN_p}+\mu_e\mathop{dN_e}+\mu_{\mu}\mathop{dN_{\mu}}.\label{eq:1st_law_extensive_npemu}
\end{align}
The Euler equation is
\begin{align}
    \varepsilon+P &= sT + \mu_nn_n+\mu_pn_p+\mu_en_e+\mu_{\mu}n_{\mu}\nonumber\\
    &= sT + \mu_nn_B-n_p\delta\mu_1+n_{\mu}\left(\delta\mu_1-\delta\mu_2\right),
\end{align}
The first law, in terms of intensive quantities, is 
\begin{equation}
    \mathop{d\varepsilon} = T\mathop{ds}+\mu_n\mathop{dn_B}-\delta\mu_1\mathop{dn_p}+\left(\delta\mu_1-\delta\mu_2\right)\mathop{dn_{\mu}}.\label{eq:1st_law_intensive_npemu}
\end{equation}
The Gibbs-Duhem equation is
\begin{equation}
    \mathop{dP}=s\mathop{dT}+n_B\mathop{d\mu_n}+(n_{\mu}-n_p)\mathop{d\delta\mu_1}-n_{\mu}\mathop{d\delta\mu_2}.\label{eq:gibbs_duhem_npemu}
\end{equation}
The first law, in terms of quantities normalized by baryon number $N_B$, is
\begin{align}
    \mathop{d\left(\frac{\varepsilon}{n_B}\right)}&=\frac{P}{n_B^2}\mathop{dn_B}+T\mathop{d\left(\frac{s}{n_B}\right)}\nonumber\\
    &-\delta\mu_1\mathop{dx_p}+\left(\delta\mu_1-\delta\mu_2\right)\mathop{dx_{\mu}}.\label{eq:first-law_per_baryon_npemu}
\end{align}

With a Legendre transformation to make temperature, not entropy per baryon, the control parameter, Eq.~\ref{eq:1st_law_intensive_npemu} and \ref{eq:first-law_per_baryon_npemu} become
\begin{align}
    \mathop{d\left(\varepsilon-sT\right)} &= \mu_n\mathop{dn_B}-s\mathop{dT}-\delta\mu_1\mathop{dn_p}\nonumber\\
    &+\left(\delta\mu_1-\delta\mu_2\right)\mathop{dn_{\mu}}\\
    \mathop{d\left(\frac{\varepsilon-sT}{n_B}\right)}&=\frac{P}{n_B^2}\mathop{dn_B}-\frac{s}{n_B}\mathop{dT}-\delta\mu_1\mathop{dx_p}\nonumber\\
    &+\left(\delta\mu_1-\delta\mu_2\right)\mathop{dx_{\mu}}.\label{eq:first_law_npemu_per_Baryon}
\end{align}

\begin{figure*}
  \centering
  \includegraphics[width=0.4\textwidth]{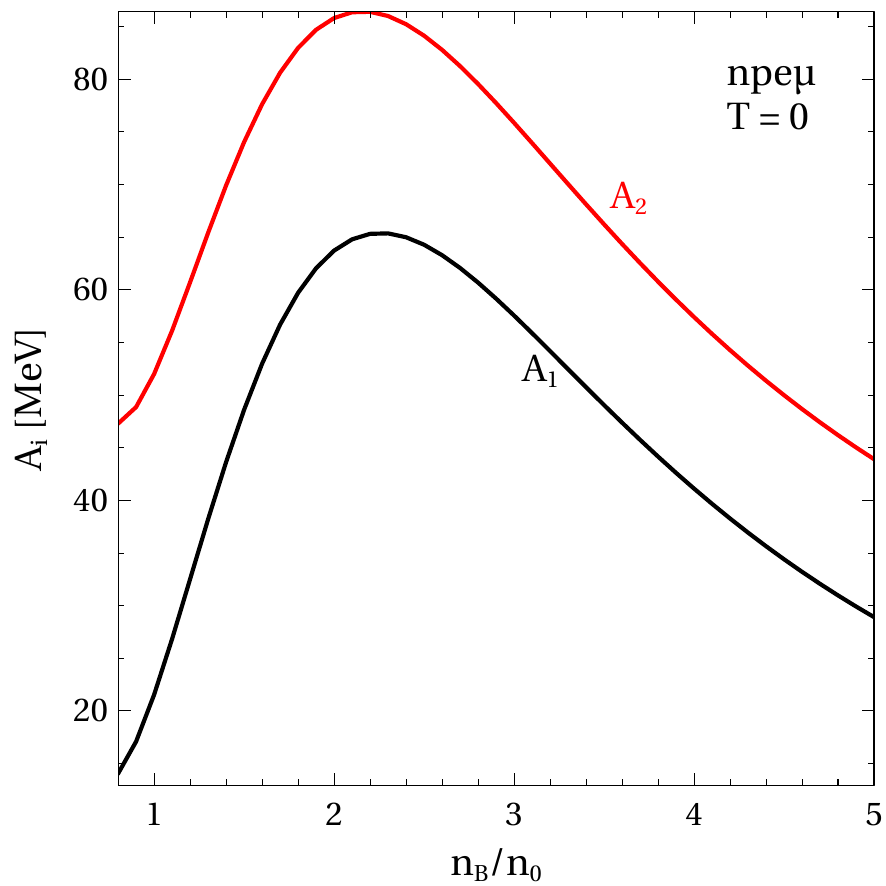}
  \includegraphics[width=0.4\textwidth]{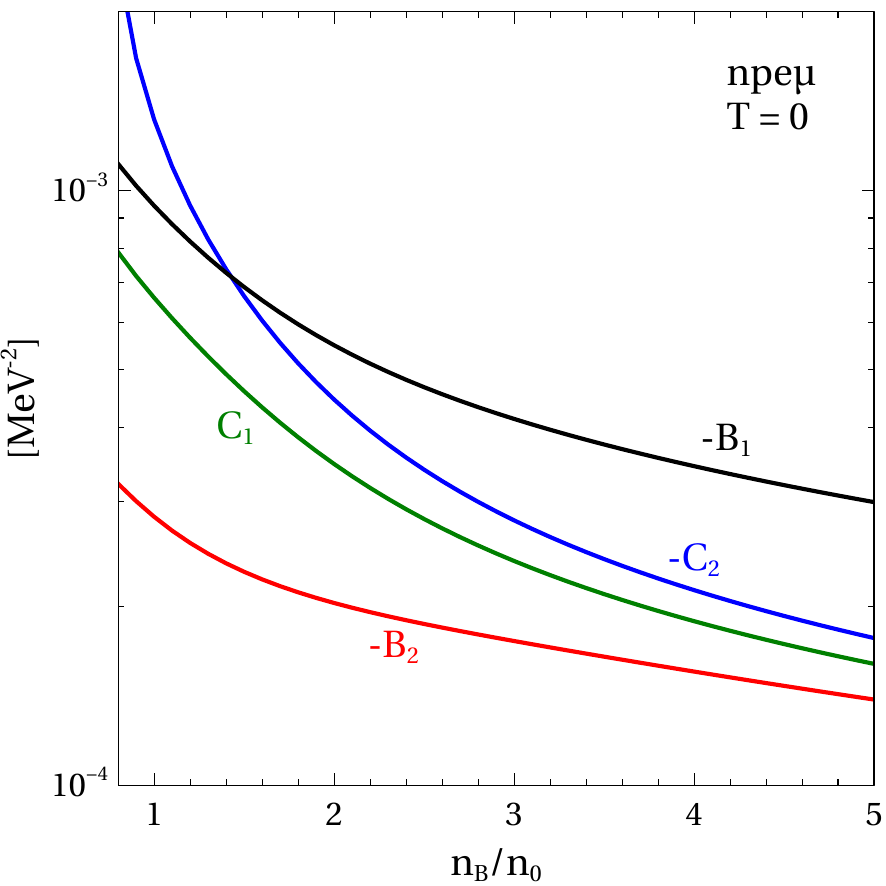}
  \caption{Isothermal susceptibilities in $npe\mu$ matter, calculated at zero temperature.  The temperature dependence is small.}
  \label{fig:A_B_C_susc_npemu}
\end{figure*}

In $npe\mu$ matter, there are many more susceptibilities.  They are defined as
\begin{subequations}
\begin{align}
    A_i &\equiv n_B \dfrac{\partial \delta\mu_i}{\partial n_B}\bigg\vert_{T,x_p,x_{\mu}},\label{eq:suscA}\\
    B_i &\equiv \dfrac{1}{n_B}\dfrac{\partial\delta\mu_i}{\partial x_p}\bigg\vert_{T,n_B,x_{\mu}},\label{eq:suscB}\\
    C_i &\equiv \dfrac{1}{n_B}\dfrac{\partial\delta\mu_i}{\partial x_{\mu}}\bigg\vert_{T,n_B,x_p}\label{eq:suscC},
\end{align}
\end{subequations}
where $i=1,2$.  
At constant temperature, three Maxwell relations can be derived from Eq.~\ref{eq:first_law_npemu_per_Baryon}
\begin{subequations}
\begin{align}
    A_1 &= -\frac{1}{n_B}\frac{\partial P}{\partial x_p}\bigg\vert_{n_B,T,x_{\mu}}\\
    A_1-A_2 &= \frac{1}{n_B}\frac{\partial P}{\partial x_{\mu}}\bigg\vert_{n_B,T,x_p}\\
    C_1 &= B_2-B_1.\label{eq:third_maxwell}
\end{align}
\end{subequations}
Eq.~\ref{eq:third_maxwell} tells us that only five of the susceptibilities, say $A_1$, $A_2$, $B_1$, $B_2$, and $C_2$ are independent.  The susceptibilities in $npe\mu$ matter are plotted in Fig.~\ref{fig:A_B_C_susc_npemu}.  Note that indeed $C_1 = B_2-B_1$
%%%%%%%%%%%%%%%%%%%%%%%%%%%%%%%%%%%%%%%%%%%%%%%%
\subsection{Bulk viscosity and beta equilibration}
The bulk viscosity of $npe\mu$ matter has been calculated in detail in neutrino-transparent matter \cite{Alford:2022ufz,Alford:2023uih} as well as in neutrino-trapped matter \cite{Alford:2021lpp,Alford:2022ufz}.  I present here a simple calculation of the $npe\mu$ bulk viscosity, where, as in the $npe$ bulk viscosity calculation, I leave out complications coming from the full phase space integration of the rates and the modification of the beta equilibrium conditions due to neutrino transparency (as described in footnote \ref{footnote:violate_beq_condition}).  Any change in the temperature throughout a density oscillation is neglected, just as in the $npe$ case.

The derivation of bulk viscosity in $npe\mu$ matter proceeds in a very similar manner to that in $npe$ matter, and the expression for the bulk viscosity (Eq.~\ref{eq:zeta_imP})
\begin{equation}
    \zeta = \left(\frac{n_B}{\delta n_B}\right)\frac{\Im{(\delta P)}}{\omega} \label{eq:bv:dp_v2}
\end{equation}
still holds.  The baryon density, pressure, and two independent particle fractions can be expanded
\begin{widetext}
\begin{subequations}
\begin{align}
    n_B(t) &= n_B + \Re{(\delta n_B e^{i\omega t})} = n_B + \delta n_B \cos{(\omega t)},\\
    P(t) &= P_0 + \Re{(\delta P e^{i\omega t})} = P_0+ \Re{(\delta P)}\cos{(\omega t)} - \Im{(\delta P)}\sin{(\omega t)},\label{eq:P_Of_t_npemu}\\
    x_p(t) &= x_p^0 + \Re{(\delta x_p e^{i\omega t})}
    = x_p^0 + \Re{(\delta x_p)}\cos{(\omega t)} - \Im{(\delta x_p)}\sin{(\omega t)},\\
    x_{\mu}(t) &= x_{\mu}^0 + \Re{(\delta x_{\mu} e^{i\omega t})}
    = x_{\mu}^0 + \Re{(\delta x_{\mu})}\cos{(\omega t)} - \Im{(\delta x_{\mu})}\sin{(\omega t)}. 
\end{align}
\end{subequations}
The pressure $P(n_B,T,x_p,x_{\mu})$ can be expanded around its equilibrium value $P_0$
\begin{align}
    P &= P_0 + \dfrac{\partial P}{\partial n_B}\bigg\vert_{T,x_p,x_{\mu}}\Re{(\delta n_B e^{i\omega t})}+\dfrac{\partial P}{\partial x_p}\bigg\vert_{T,n_B,x_{\mu}}\Re{(\delta x_p e^{i\omega t})}+\dfrac{\partial P}{\partial x_{\mu}}\bigg\vert_{T,n_B,x_{p}}\Re{(\delta x_{\mu} e^{i\omega t})},\label{eq:P_npemu}\\
    &= P_0 + \frac{1}{\kappa_T}\frac{\delta n_B}{n_B}\cos{(\omega t)}-n_BA_1\left[\Re{(\delta x_p)}\cos{(\omega t)}-\Im{(\delta x_p)}\sin{(\omega t)}\right]+n_B(A_1-A_2)\left[\Re{(\delta x_{\mu})}\cos{(\omega t)}-\Im{(\delta x_{\mu})}\sin{(\omega t)}\right].\nonumber
\end{align}
\end{widetext}
Matching sine and cosine terms of Eqs.~\ref{eq:P_Of_t_npemu} and \ref{eq:P_npemu} yields
\begin{subequations}
\begin{align}
    \Re{(\delta P)} &= \frac{1}{\kappa_T}\frac{\delta n_B}{n_B}-n_BA_1\Re{(\delta x_p)}\nonumber\\
    &+n_B(A_1-A_2)\Re{(\delta x_{\mu})}\\
    \Im{(\delta P)} &= -n_BA_1\Im{(\delta x_p)}+n_B(A_1-A_2)\Im{(\delta x_{\mu})}.\label{eq:imdP_npemu}
\end{align}
\end{subequations}

\begin{table*}[]
\begin{tabular}{l|l|l}\hline
\multicolumn{1}{|c|}{Equilibrating $\delta\mu$} & \multicolumn{1}{c|}{Reactions} & \multicolumn{1}{c|}{$\overrightarrow{\Gamma} - \overleftarrow{\Gamma}$ (subthermal)} \\ \hline
\multicolumn{1}{|l|}{\multirow{2}{*}{1) $\delta\mu_1 \equiv \mu_n-\mu_p-\mu_e$}} & $n \rightarrow p + e^- + \bar{\nu}_e$ & \multicolumn{1}{l|}{\multirow{2}{*}{$\lambda_{1} \delta\mu_1$}} \\
\multicolumn{1}{|l|}{} & $e^- + p \rightarrow n + \nu_e$ & \multicolumn{1}{l|}{} \\ \hline
\multicolumn{1}{|l|}{\multirow{2}{*}{2) $\delta\mu_2\equiv \mu_n - \mu_p -\mu_{\mu}$}} & $n \rightarrow p + \mu^- + \bar{\nu}_{\mu}$ & \multicolumn{1}{l|}{\multirow{2}{*}{$\lambda_{2} \delta\mu_2$}} \\
\multicolumn{1}{|l|}{} & $\mu^- + p \rightarrow n + \nu_{\mu}$ & \multicolumn{1}{l|}{} \\ \hline
\multicolumn{1}{|l|}{\multirow{2}{*}{3) $\delta\mu_3 = \mu_{\mu}-\mu_e = \delta\mu_1 - \delta\mu_2$}} & $\mu^-\rightarrow e^- + \nu_{\mu}+\bar{\nu}_e$ & \multicolumn{1}{l|}{\multirow{2}{*}{$\lambda_{\mu e} (\delta\mu_1-\delta\mu_2)$}} \\
\multicolumn{1}{|l|}{} & $e^- \rightarrow \mu^- + \nu_e + \bar{\nu}_{\mu}$ & \multicolumn{1}{l|}{} \\ \hline
\end{tabular}
\label{table:npemu_rxns}
\caption{Chemical equilibration mechanisms at work in $npe\mu$ matter.  While the ``direct'' versions of the chemical processes are written in the middle column, the modified versions of the processes (i.e., those with spectator particles) operate too, and in the conditions discussed in this chapter, are the principle operating processes because the direct processes are kinematically forbidden (though see footnote \ref{footnote:muonconversion}).}
\end{table*}

To solve for the particle fractions, we need to study the flavor-changing interactions in the $npe\mu$ system.  The possible reactions are summarized in Table I.  The two independent particle fractions $x_p$ and $x_{\mu}$ evolve according to the equations (jumping immediately to the subthermal limit)
\begin{subequations}
\begin{align}
    n_B\frac{\mathop{dx_p}}{\mathop{dt}}&=\lambda_{1}\delta\mu_1+\lambda_{2}\delta\mu_2,\label{eq:dxpdt_npemu}\\
    n_B\frac{\mathop{dx_{\mu}}}{\mathop{dt}} &= -\lambda_{\mu e}\delta\mu_1+(\lambda_{2}+\lambda_{\mu e})\delta\mu_2.\label{eq:dxmudt_npemu}
\end{align}
\end{subequations}
\begin{figure*}
  \centering
  \includegraphics[width=0.4\textwidth]{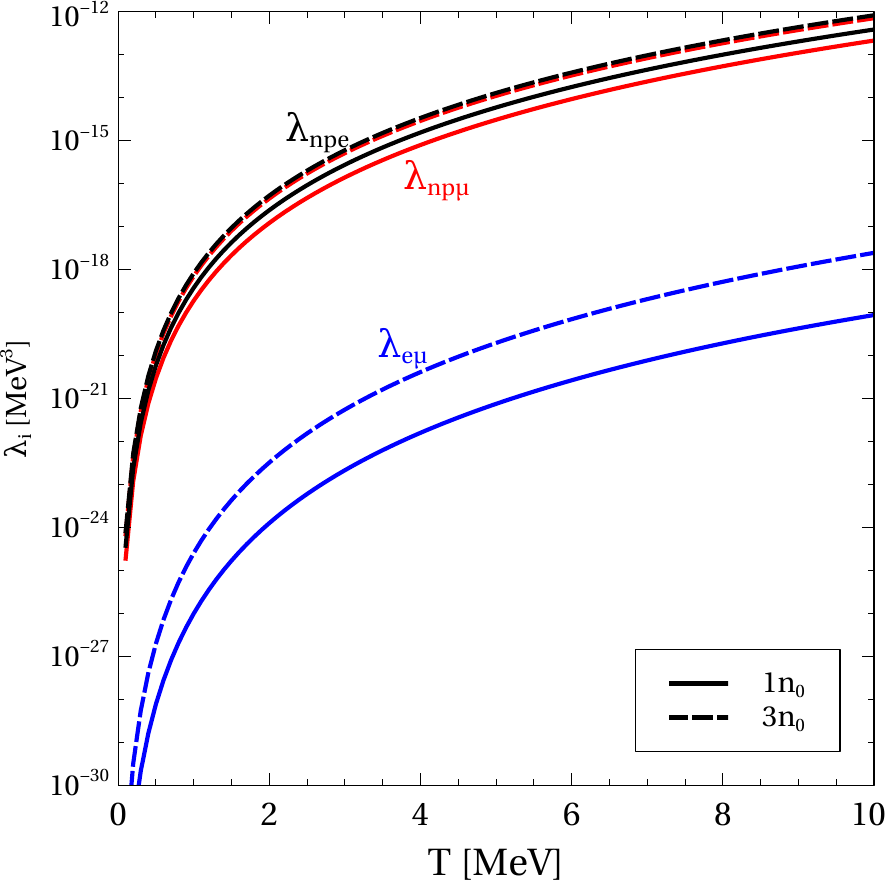}
  \includegraphics[width=0.4\textwidth]{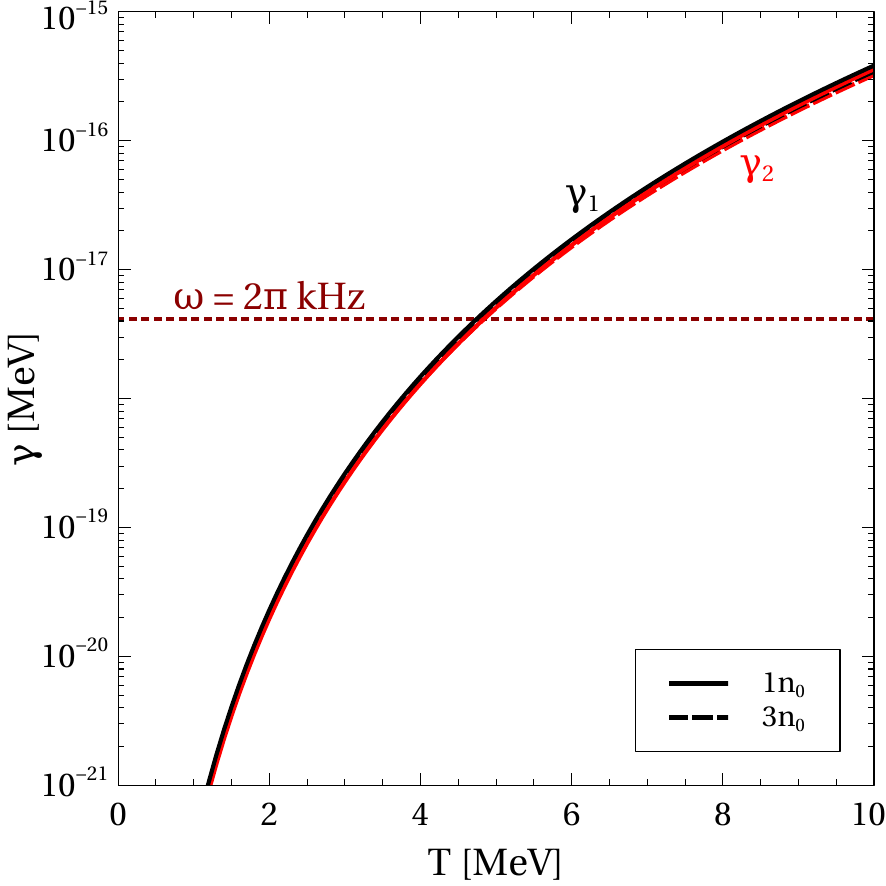}
  \caption{Left: Beta reaction rates $\lambda$ versus temperature ($\lambda_{npe}$ is $\lambda_1$ and $\lambda_{np\mu}$ is $\lambda_2$).  Right: Beta equilibration rates $\gamma$ of the partial bulk viscosities.  They are defined as $\gamma_1\equiv \vert B_1\vert\lambda_1$ and $\gamma_2\equiv \vert B_2+C_2\vert\lambda_2$ and should be compared to the density oscillation frequency, which in neutron star mergers is typically around 1 kHz ($\omega = 2\pi\text{ kHz}$).}
  \label{fig:lambda_npemu}
\end{figure*}
The subthermal rates $\lambda_i$ are 
\begin{subequations}
    \begin{align}
        \lambda_{1} &= \frac{367}{1152\pi^3}G^2g^2_Af^4\frac{m_n^4}{m^4_{\pi}}\frac{p^4_{Fn}p_{Fp}}{(p^2_{Fn}+m^2_{\pi})^2}T^6\theta_n.\label{eq:lambda_npe_2}\\
        \lambda_{2} &= \frac{367}{1152\pi^3}G^2g^2_Af^4\frac{m_n^4}{m^4_{\pi}}\frac{p^4_{Fn}p_{Fp}}{(p^2_{Fn}+m^2_{\pi})^2}\frac{p_{F\mu}}{\sqrt{p_{F\mu}^2+m_{\mu}^2}}T^6\theta_n.\label{eq:lambda_npemu_2}
    \end{align}
\end{subequations}
The expression\footnote{In neutrino-transparent matter, the process $\mu^-\rightarrow e^-+\nu_{\mu}+\bar{\nu}_{e}$ is actually allowed, but the reverse process $e^-\rightarrow \mu^-+\bar{\nu}_{\mu}+\nu_{e}$ is not, which could complicate the analysis here \cite{Alford:2023uih}. 
 I neglect both of these ``direct'' processes in this chapter.\label{footnote:muonconversion}} for $\lambda_{\mu e}$ is given in terms of a convenient curve fit in Eqs.~32-35 in Alford \& Good \cite{Alford:2010jf}.  In the expression for $\lambda_2$, the term $\theta_n$ is defined in Eq.~\ref{eq:theta_n}, but with $p_{Fe}$ replaced by $p_{F\mu}$.  The functions $\lambda_i$ are plotted in the left panel of Fig.~\ref{fig:lambda_npemu} in matter described by the IUF EoS.

The deviations from chemical equilibrium can be expanded around equilibrium and written in terms of the susceptibilities
\begin{subequations}
\begin{align}
    \delta\mu_1 &= A_1\frac{\delta n_B}{n_B}\cos{(\omega t)}\nonumber\\
    &+n_BB_1\left[\Re{(\delta x_p)}\cos{(\omega t)}-\Im{(\delta x_p)}\sin{(\omega t)}\right]\\
    &+n_BC_1\left[\Re{(\delta x_{\mu})}\cos{(\omega t)}-\Im{(\delta x_{\mu})}\sin{(\omega t)}\right],\nonumber\\
    \delta\mu_2 &= A_2\frac{\delta n_B}{n_B}\cos{(\omega t)}\nonumber\\
    &+n_BB_2\left[\Re{(\delta x_p)}\cos{(\omega t)}-\Im{(\delta x_p)}\sin{(\omega t)}\right]\\
    &+n_BC_2\left[\Re{(\delta x_{\mu})}\cos{(\omega t)}-\Im{(\delta x_{\mu})}\sin{(\omega t)}\right].\nonumber
\end{align}
\end{subequations}
In the same manner as in the previous section, Eqs.~\ref{eq:dxpdt_npemu} and \ref{eq:dxmudt_npemu} lead to a system of equations for the real and imaginary parts of $\delta x_p$ and $\delta x_{\mu}$.  The solutions are proportional to $\delta n_B/n_B$ and are functions of the five independent susceptibilities, the three rates $\lambda_i$, and the density oscillation frequency $\omega$, but are too complicated to be shown here.  The bulk viscosity is obtained with Eqs.~\ref{eq:bv:dp_v2} and \ref{eq:imdP_npemu}, and is given by
\begin{equation}
    \zeta_{npe\mu} = \dfrac{F+G\omega^2}{H+J\omega^2+\omega^4},\label{eq:bv_npemu_general_expression}
\end{equation}
with
\begin{widetext}
\begin{subequations}
\begin{align}
    &F = \left[\lambda_{2}\lambda_{\mu e}+\lambda_{1}(\lambda_{2}+\lambda_{\mu e})\right]\left\{(A_1B_2-A_2B_1)^2\lambda_{1}+\left[(A_1-A_2)B_2+A_1C_2\right]^2\lambda_{2}+\left[A_2(B_1-B_2)+A_1C_2\right]^2\lambda_{\mu e}\right\},\\
    &G = A_1^2\lambda_{1}+A_2^2\lambda_{2}+(A_1-A_2)^2\lambda_{\mu e},\\
    &H = \left[\lambda_{2}\lambda_{\mu e}+\lambda_{1}(\lambda_{2}+\lambda_{\mu e})\right]^2\left[B_2^2-B_1(B_2+C_2)\right]^2,\\
    &J = B_1^2\lambda_{1}^2+2B_2^2\lambda_{1}\lambda_{2}+(B_2+C_2)^2\lambda_{2}^2+2(B_1-B_2)^2\lambda_{1}\lambda_{\mu e}+2C_2^2\lambda_{2}\lambda_{\mu e}+(B_2-B_1-C_2)^2\lambda_{\mu e}^2.
\end{align}
\end{subequations}
\end{widetext}
This form of the bulk viscosity, $\zeta_{npe\mu} = (F+G\omega^2)/(H+J\omega^2+\omega^4)$, is present in other systems with two equilibrating quantities, such as neutrino-trapped $npe\mu$ matter \cite{Alford:2021lpp,Alford:2022ufz} and strange quark matter phases \cite{Sad:2007afd,Alford:2006gy,Wang:2010ydb}.

Figure \ref{fig:lambda_npemu} indicates that the rate $\lambda_{\mu e}$ is much slower than the other rates.  In fact, setting $\lambda_{\mu e}=0$ in Eq.~\ref{eq:bv_npemu_general_expression} does not affect the bulk viscosity at all in the conditions studied here.  The bulk viscosity with $\lambda_{\mu e}=0$ is
\begin{widetext}
\begin{equation}
     \zeta_{npe\mu} = \dfrac{\lambda_{1}\lambda_{2}\left\{(A_1B_2-A_2B_1)^2\lambda_{1}+\left[(A_1-A_2)B_2+A_1C_2\right]^2\lambda_{2}\right\}+(A_1^2\lambda_{1}+A_2^2\lambda_{2})\omega^2}{\lambda_{1}^2\lambda_{2}^2\left[B_2^2-B_1(B_2+C_2)\right]^2+\left[B_1^2\lambda_{1}^2+2B_2^2\lambda_{1}\lambda_{2}+(B_2+C_2)^2\lambda_{2}^2\right]\omega^2+\omega^4},\label{eq:bv_npemu_l6_is_zero}
\end{equation}
\end{widetext}
which of course still has its $\zeta_{npe\mu} = (F+G\omega^2)/(H+J\omega^2+\omega^4)$ form (with different $F,G,H,J$) because there are still two equilibrating quantities (c.f.~Eq.~\ref{eq:dxmudt_npemu} with $\lambda_{\mu e}=0$).
In interpreting Eq.~\ref{eq:bv_npemu_l6_is_zero}, it is useful to define the ``partial'' bulk viscosities - the bulk viscosity due to each reaction $\lambda_i$, with $\lambda_{j\neq i}=0$.  We have
\begin{subequations}
\begin{align}
\zeta_{1} &\equiv \frac{A_1^2\lambda_{1}}{B_1^2\lambda_{1}^2+\omega^2} = \frac{A_1^2}{\vert B_1\vert}\frac{\gamma_1}{\gamma_1^2+\omega^2}\label{eq:bv_npemu_l1only}\\
\zeta_{2} &\equiv \frac{A_2^2\lambda_{2}}{(B_2+C_2)^2\lambda_{2}^2+\omega^2}=\frac{A_2^2}{\vert B_2+C_2\vert}\frac{\gamma_2}{\gamma_2^2+\omega^2},\label{eq:bv_npemu_l2only}
\end{align}
\end{subequations}
where in analogy with Eq.~\ref{eq:gamma_npe}, the beta equilibration rates $\gamma_i$ in each partial bulk viscosity are defined
\begin{subequations}
    \begin{align}
        \gamma_1 &\equiv \vert B_1\vert\lambda_1,\\
        \gamma_2 &\equiv \vert B_2+C_2\vert \lambda_2.  
    \end{align}
\end{subequations}
These equilibration rates $\gamma_i$ are plotted in the right panel of Fig.~\ref{fig:lambda_npemu}.  The two rates $\gamma_1$ and $\gamma_2$ are nearly equal across the range of temperatures and densities shown in the plot, and they match the 1 kHz frequency of the density oscillation at $T\approx 4-5\text{ MeV}$. 

Eq.~\ref{eq:bv_npemu_l1only} is just the bulk viscosity in $npe$ matter equilibrating by the Urca process (with electrons) described by Eq.~\ref{eq:bv_npe}.  In the high-frequency limit the bulk viscosity (Eq.~\ref{eq:bv_npemu_l6_is_zero}) expression decouples
\begin{equation}
    \zeta_{npe\mu}^{\gamma_i\ll\omega} = \frac{A_1^2\lambda_1+A_2^2\lambda_2}{\omega^2}=\zeta_1^{\gamma\ll\omega}+\zeta_2^{\gamma\ll\omega}
\end{equation}
into a sum of two individual contributions.  The high-frequency limit is applicable in cold neutron stars, which explains why many of the early bulk viscosity papers, which focused on cold neutron stars, considered the total bulk viscosity of a multicomponent system to be a sum of all of the individual contributions (e.g.~\cite{Haensel:2000vz,Haensel:2001mw,Haensel:2001em}).  

\begin{figure*}
  \centering
  \includegraphics[width=0.4\textwidth]{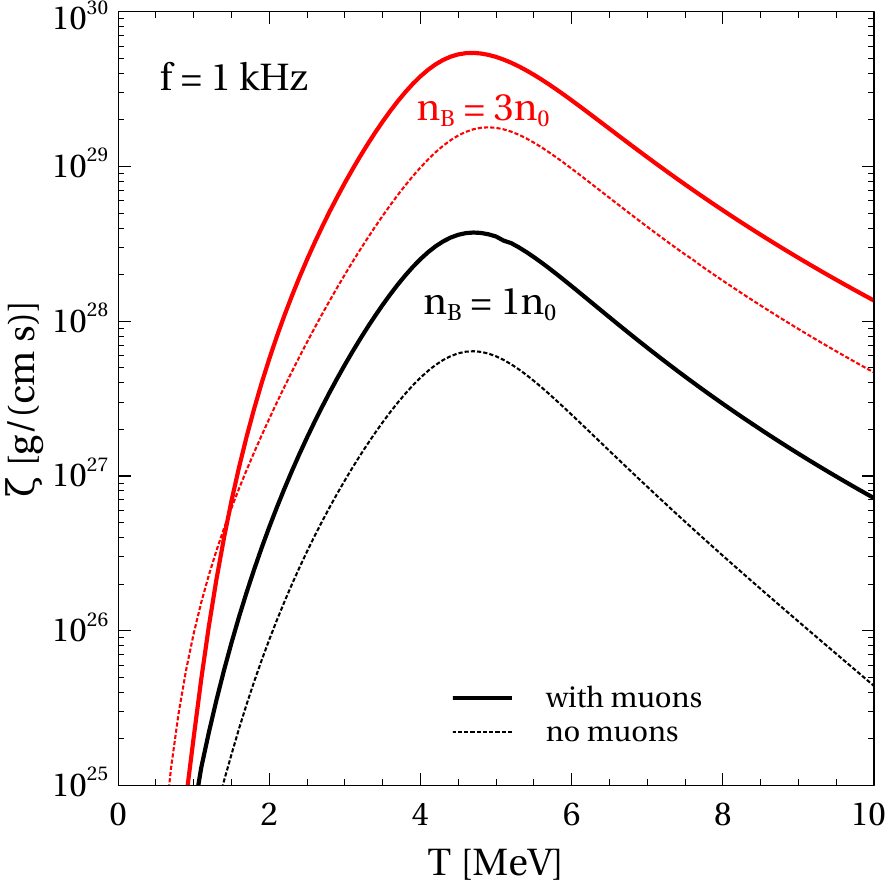}
  \includegraphics[width=0.4\textwidth]{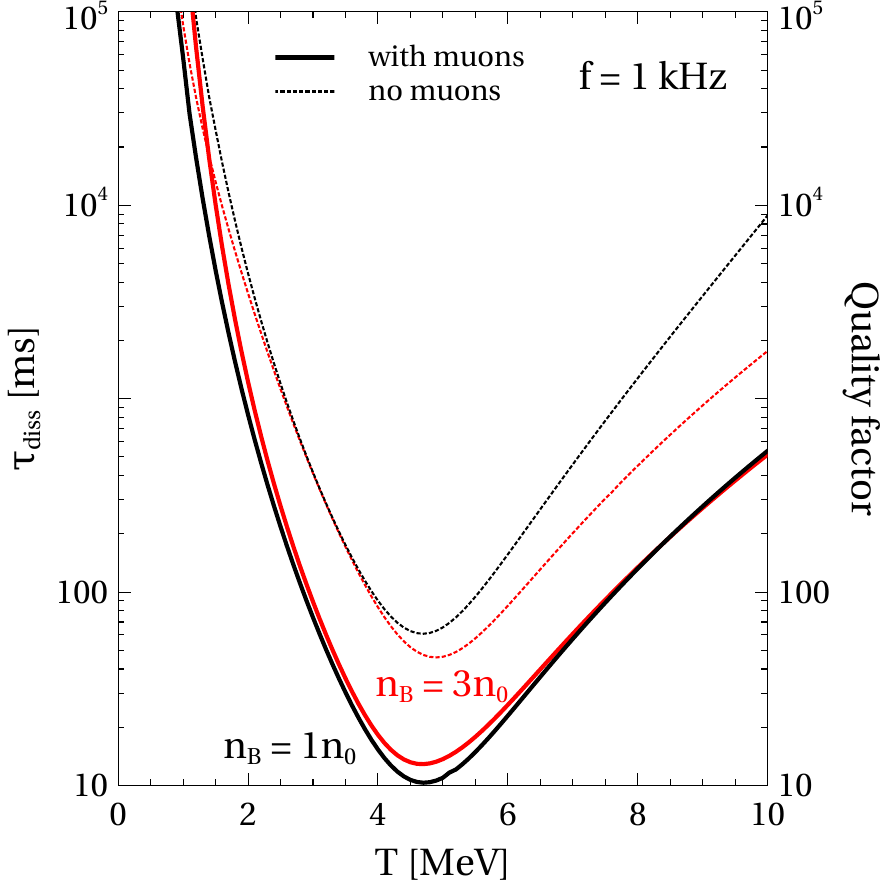}
  \caption{Left: Bulk viscosity in $npe\mu$ matter subjected to a harmonic, small-amplitude density oscillation of angular frequency $\omega = 2\pi\text{ kHz}.$  Right: The timescale of energy dissipation of the oscillation due to the bulk viscosity of $npe\mu$ matter, as well as the quality factor of the oscillation.}
  \label{fig:bulk_viscosity_npemu}
\end{figure*}

The bulk viscosity of $npe\mu$ matter undergoing a 1 kHz density oscillation is plotted in the left panel of Fig.~\ref{fig:bulk_viscosity_npemu}.  The dashed lines show the results without muons, and the solid lines with muons included (both in the EoS and in equilibration processes).  There is still essentially one resonance, which at first thought is surprising, since there are two equilibrating quantities, but this behavior will be explained shortly.  The presence of muons is found to increase the bulk viscosity by about a factor of 2-5 at the resonant maximum.  The resonance occurs roughly at the same temperature, because adding muons to the system does not significantly change the equilibration rates in the $npe$ sector, and the equilibration rates in the $np\mu$ sector are very close to the $npe$ rates (see the right panel of Fig.~\ref{fig:lambda_npemu}).  The dissipation timescale is shown in the right panel of Fig.~\ref{fig:bulk_viscosity_npemu}.  Matter with muons has a strong enough bulk viscosity to damp density oscillations in less than 10 ms.  The quality factor of kHz oscillations in $npe\mu$ matter can be as low as 10.

\begin{figure}
  \centering
  \includegraphics[width=0.4\textwidth]{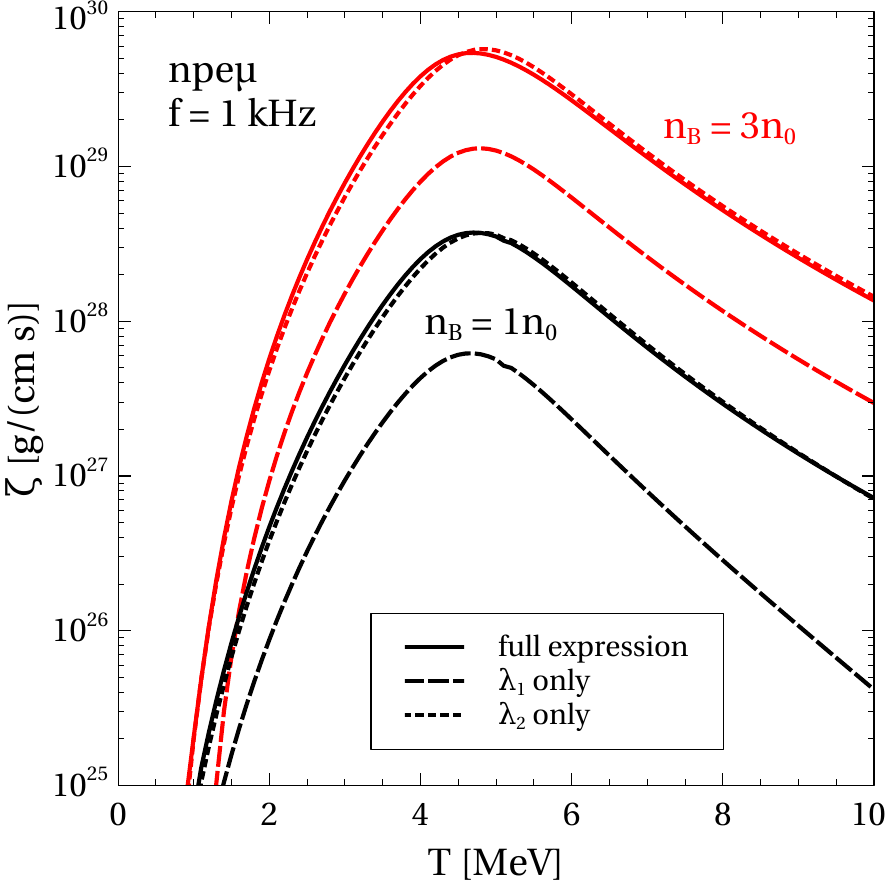}
  \caption{Bulk viscosity in $npe\mu$ matter, compared with the partial bulk viscosities $\zeta_1$ and $\zeta_2$.}
  \label{fig:npemu_partial_bv}
\end{figure}

To further understand the behavior of the bulk viscosity curve, Fig.~\ref{fig:npemu_partial_bv} shows the bulk viscosity in $npe\mu$ matter and also the partial bulk viscosities $\zeta_1$ and $\zeta_2$ ($\zeta_{\mu e}$, if defined in an analogous way to the other partial bulk viscosities, is much smaller than $\zeta_1$ and $\zeta_2$).  From this plot, it is clear that the total bulk viscosity is \textit{not} the sum of the partial bulk viscosities (except in the high-frequency limit, which corresponds to temperatures below $\approx 4 \text{ MeV}$ on this plot).  The right panel of Fig.~\ref{fig:lambda_npemu} indicates that the resonances of the partial bulk viscosities $\zeta_1$ and $\zeta_2$ both lie near $T=4-5\text{ MeV}$.  In the FS approximation used here for the rates, the electron mUrca rate is nearly identical to the muon mUrca rate, because the mUrca rates depend on the lepton type by just a slowly varying function of the lepton density (Eq.~\ref{eq:lambda_npe_2} and \ref{eq:lambda_npemu_2}).  In reality, the below-threshold contribution of the direct Urca rates, which is significant for $T\gtrsim 1\text{ MeV}$ \cite{Alford:2018lhf,Alford:2021ogv}, would split the near degeneracy between the rates \cite{Alford:2023uih}.  

The behavior of the total bulk viscosity near the two overlapping resonances of $\zeta_1$ and $\zeta_2$ is difficult to predict because the resonances interact in a complicated way.  The height of the resonance of $\zeta_2$ is larger than $\zeta_1$ because the susceptibility combination $A_2^2/\vert B_2+C_2\vert$ is several times larger than $A_1^2/\vert B_1\vert$. 
 Because of this dominance of $\zeta_2$, the total bulk viscosity follows $\zeta_2$ quite closely, and the peak of the curve shifts only slightly to the left, representing the influence of the resonance of $\zeta_1$.

\begin{figure*}
  \centering
  \includegraphics[width=0.4\textwidth]{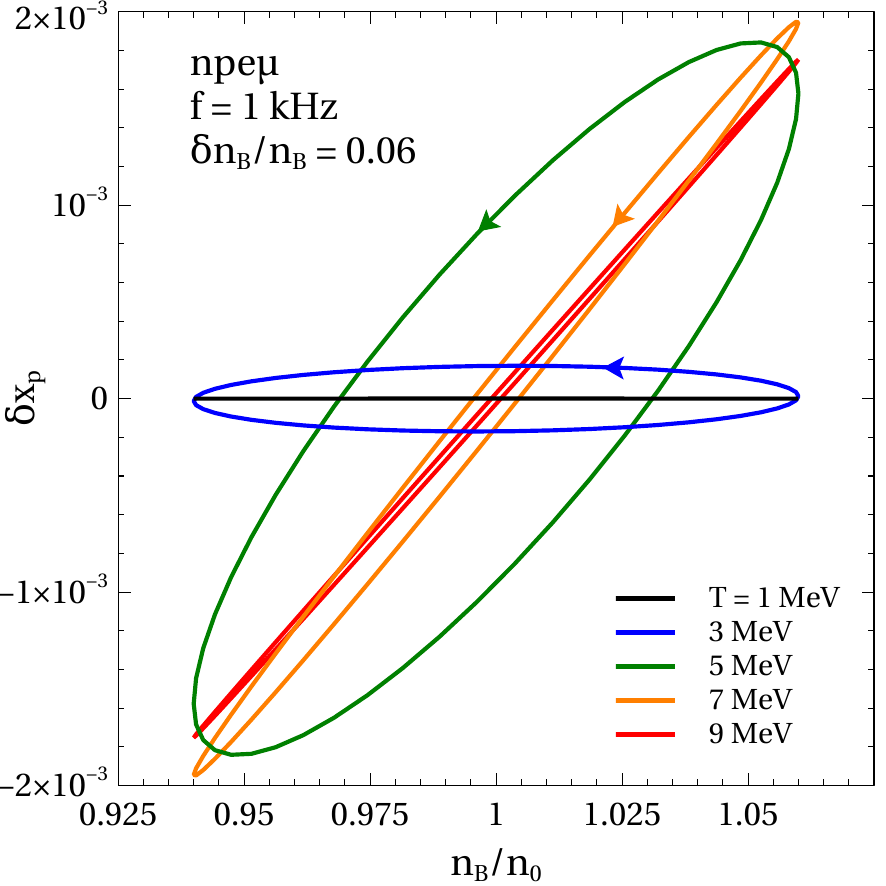}
  \includegraphics[width=0.4\textwidth]{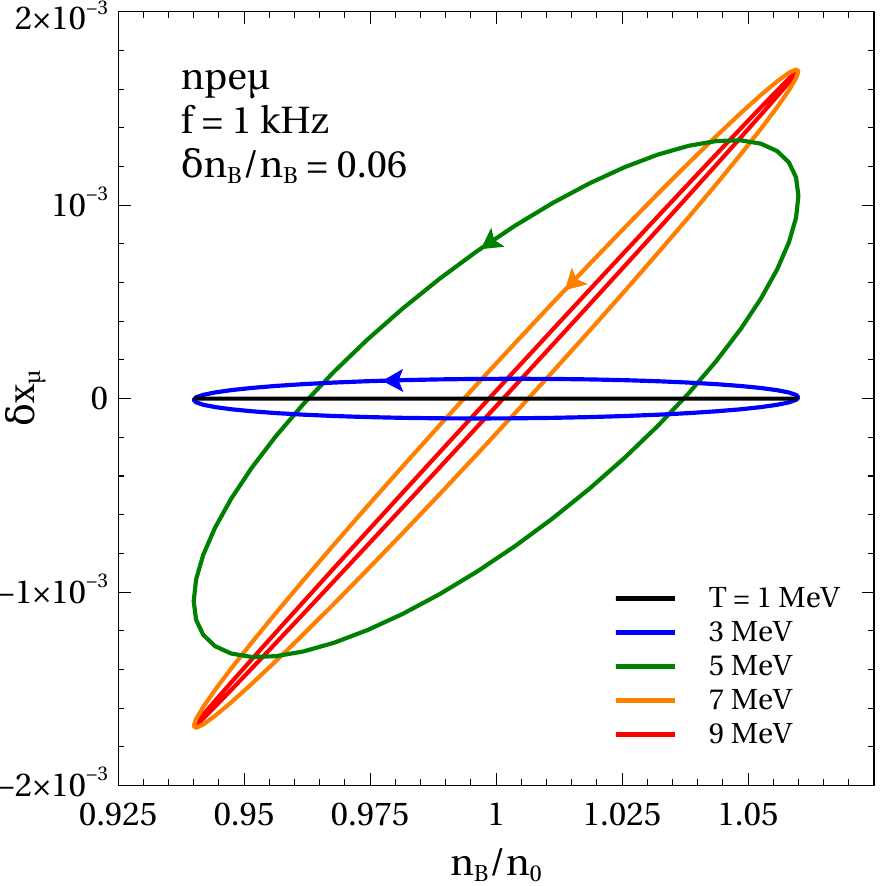}
  \caption{Path of a fluid element in the $x_pn_B$ plane (left) and the $x_{\mu}n_B$ plane (right), over the course of a complete period of the density oscillation.  Five trajectories, each at a different temperature (leading to different beta equilibration rates) are shown.}
  \label{fig:xp_and_xmu_vs_nB_npemu}
\end{figure*}

In Fig.~\ref{fig:xp_and_xmu_vs_nB_npemu} is shown the path of a fluid element in the $x_pn_B$ (left panel) and $x_{\mu}n_B$ (right panel) planes over the course of a full oscillation period.  The behavior is very much like that observed in the $npe$ system.  At low temperature, the equilibration rates are very slow and the particle fractions remain close to unchanged throughout the oscillation.  As the temperature increases above 1 MeV, the Urca reactions are fast enough to make progress toward establishing beta equilibrium, but the particle fractions still lag behind the density change, creating a curve with finite area in the $x_in_B$ plane.  The area, and thus the dissipation, is maximal at temperatures near 5 MeV, and when the temperature exceeds that, the Urca processes are fast enough to keep the system close to beta equilibrium throughout the oscillation.

\begin{figure*}
  \centering
  \includegraphics[width=0.4\textwidth]{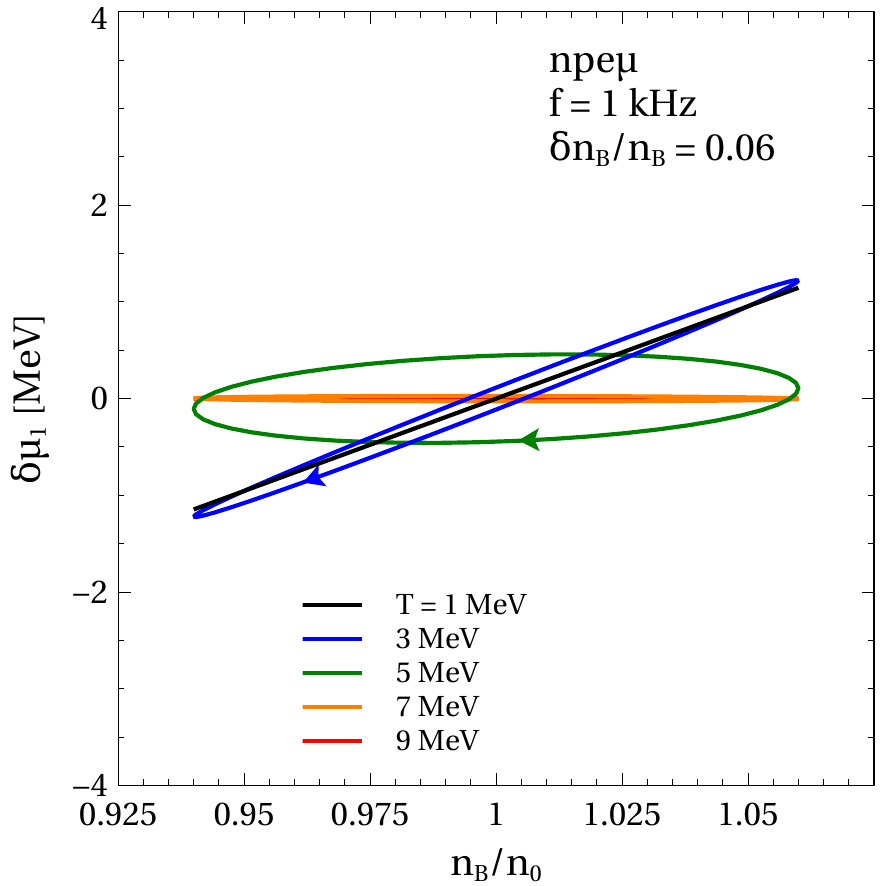}
  \includegraphics[width=0.4\textwidth]{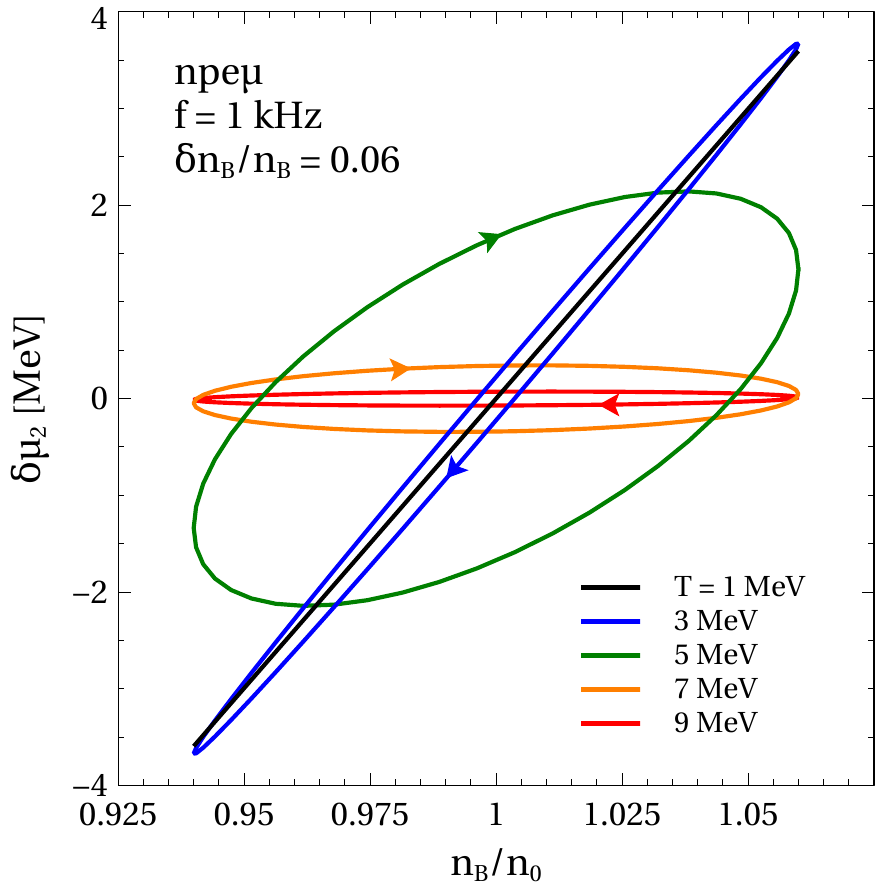}
  \caption{Path of a fluid element in the $\delta\mu_1n_B$ plane (left) and the $\delta\mu_2n_B$ plane (right) over the course of a complete period of the density oscillation.  Five trajectories, each at a different temperature (leading to different beta equilibration rates) are shown.}
  \label{fig:dmu_npemu}
\end{figure*}

Fig.~\ref{fig:dmu_npemu}, which plots the path of a fluid element in the $\delta\mu_1n_B$ (left panel) and $\delta\mu_2nB$ plans, shows very similar information.  When the system is at high temperature and is efficiently chemically equilibrated, $\delta\mu$ deviates little from zero.  When the temperature is low and the system struggles to chemically equilibrate, $\delta\mu$ is able to grow large.  In fact, for temperatures of 1 and 3 MeV, at least one of the $\delta\mu$ values exceeds the temperature, and thus the system is no longer in the subthermal limit and the results in the plot cannot be trusted.
%%%%%%%%%%%%%%%%%%%%%%%%%%%%%%%%%%%%%%%%%%%%%%%%%%%%%
\section{Neutrino-trapped nuclear matter}\label{sec:nu_trapped}
The neutrino MFP in dense matter decreases significantly as the temperature increases.  At temperatures above about 5 MeV, the MFP\footnote{The neutrino MFP is actually a function of neutrino energy.  It is common practice to, at a particular density and temperature, calculate the MFP of a ``typical'' neutrino with energy $E_{\nu}\approx 3T$ (the average energy of massless fermions with a Fermi-Dirac distribution at zero chemical potential is $\langle E\rangle = 7\pi^4T/\left[180\zeta (3)\right]\approx 3.15T$) and call that \textit{the} neutrino MFP.} falls below a kilometer \cite{Alford:2018lhf,Roberts:2016mwj} and neutrinos can be considered trapped inside a ten-kilometer-scale object like a neutron star.  In this case, the nuclear matter picks up a new conserved quantity, the total lepton number $Y_L\equiv (n_e+n_{\nu})/n_B$.  The value of the conserved lepton fraction $Y_L$ is determined by the history of the system, in the same way that the conserved baryon number $N_B$ is.  A neutron star in isolation produces neutrinos, but they escape easily due to the low temperature.  When two neutron stars merge, much of the matter is heated to temperatures of tens of MeV, trapping any neutrinos produced afterwords.  Whatever the lepton number is in a fluid element at the point when it begins to trap neutrinos becomes conserved.  

Matter with trapped neutrinos (but no muons) only has one independent particle fraction, say $x_p$, because the addition of neutrinos comes with the addition of a constraint (conservation of lepton number). 
 Neutrino-trapped nuclear matter, let's say with a net electron neutrino population (as opposed to antineutrinos), achieves chemical equilibrium through the direct Urca electron capture processes
\begin{equation}
    n+\nu \leftrightarrow e^- + p.
\end{equation}
Neutron decay and its inverse, as well as modified Urca processes are slower and can be neglected in the case of matter with a net neutrino population.  In the limit of strongly degenerate matter (note that this means a substantial neutrino population with Fermi momentum $p_{F\nu}$), the rate of the direct Urca electron capture process in the subthermal limit is \cite{Alford:2020lla}
\begin{widetext}
\begin{equation}
    \lambda_{\text{dUrca, trapped}} = \frac{1}{12\pi^3}G_F^2\cos^2{\theta_C}(1+3g_A^2)E_{Fn}^*E_{Fp}^*p_{Fe}p_{F\nu}(p_{Fp}+p_{Fe}+p_{F\nu}-p_{Fn})T^2.
\end{equation}
This rate is very fast, and has no kinematic threshold.  For comparison, the subthermal direct Urca rate in neutrino-transparent nuclear matter is \cite{Alford:2020lla,Alford:2019qtm}
\begin{equation}
    \lambda_{\text{dUrca, transparent}} = \frac{17}{240\pi}G_F^2\cos^2{\theta_C}(1+3g_A^2)E_{Fn}^*E_{Fp}^*p_{Fe}T^4\theta_{\text{dUrca}},
\end{equation}
where $\theta_{\text{dUrca}}$ is one if $p_{Fn}<p_{Fp}+p_{Fe}$ and zero otherwise.  The ratio of these two rates (when $\theta_{\text{dUrca}}=1$) is
\begin{equation}
    \frac{\lambda_{\text{dUrca, trapped}}}{\lambda_{\text{dUrca, transparent}}} = \frac{20}{17\pi^2}\frac{p_{F\nu}(p_{Fp}+p_{Fe}+p_{F\nu}-p_{Fn})}{T^2},
\end{equation}
which is typically much larger than one.\end{widetext}
Therefore, neutrino-trapped matter equilibrates much faster than even dUrca-equilibrated neutrino-transparent matter and therefore, assuming the susceptibilities are not changed dramatically with the addition of neutrinos\footnote{Independent of the issue of neutrino-trapped versus neutrino-transparent susceptibilities, at temperatures of tens of MeV, the isothermal and adiabatic susceptibilities differ significantly from each other \cite{Benhar:2023mgk,Alford:2022ufz}.}, the bulk-viscous resonance likely occurs at a temperature well below 1 MeV.  In fact, calculations by Alford, Harutyunyan, \& Sedrakian \cite{Alford:2021lpp} confirm that this is true.  Their calculation includes all relevant neutrino and antineutrino processes, and the rate calculations go beyond the FS approximation by doing the full phase space integration.  The neutrino-trapped regime is $T\gtrsim 5\text{ MeV}$, and so the bulk viscosity is very far from the resonant peak. 
 The calculations in \cite{Alford:2021lpp} indicate that the bulk viscosity in neutrino-trapped matter is only strong enough to damp density oscillations on the timescale of seconds or longer.
%%%%%%%%%%%%%%%%%%%%%%%%%%%%%%%%%%%%%%%%%%%%%%%%
\section{Effects of bulk viscosity in neutron star mergers}\label{sec:bv_effects}
Neutron star mergers, both in the inspiral and postmerger phase, 
offer an interesting environment to search for signatures of bulk-viscous dissipation because not only is the matter likely to experience changes in density (ranging from small amplitude oscillations to catastrophic changes in density during a potential collapse to a black hole), but because we can see both gravitational and electromagnetic signals from neutron star mergers, giving the community many tools with which to search for dissipative effects.
%%%%%%%%%%%%%%%%%%%%%%%%%%%%%%%%%%%%%%
\subsection{Neutron star inspiral}\label{sec:inspiral}
In a typical neutron star binary, the stars orbit around their common center of mass for millions of years, emitting gravitational radiation that takes energy away from their orbit, causing them to grow closer together over time \cite{Andersson:2019yve}.  Eventually, the two stars get close enough to each other that they no longer are effectively point objects, and tidal forces become significant.  By the time the neutron stars are close enough to tidally interact, they (typically) have been cooling via neutrino and photon emission for long enough to have core temperatures around $10^6-10^7\text{ K}$ \cite{Potekhin:2017ufy}.  

Arras \& Weinberg \cite{Arras:2018fxj} studied the role that weak interactions play in dissipating energy in tidally interacting, inspiraling neutron stars.  Tidal interactions excite oscillation modes within a star, pushing it out of beta equilibrium.  The matter can deviate significantly from beta equilibrium because the Urca rates in $T\ll 1\text{ MeV}$ matter are much slower than the dynamical timescale of milliseconds.  Nevertheless, in response, the Urca processes turn on, attempting to reestablish beta equilibrium, and the neutrino emission is enhanced.  In these cold neutron stars, even a very small amplitude density oscillation (say, $\delta n_B/n_B \sim 0.01$) is not subthermal (this should seem plausible from examining Fig.~\ref{fig:dmu_over_t}, though the smallest temperature considered in the plot is around $10^9$ K), and therefore one must use the full expression for the Urca rates out of beta equilibrium (e.g.~Eq.~\ref{eq:suprathermal_murca} or see the appendix of \cite{Most:2022yhe}).  As described in Sec.~\ref{sec:npe_bv} and \ref{sec:npemu_bv} in this chapter, the chemical equilibration processes result in $PdV$ work done on the fluid element, leading to heating.  In the subthermal limit, the extra cooling from the enhanced neutrino emission wins over the heating, and the fluid element still cools, though more slowly than if it were not vibrating.  Suprathermal oscillations can actually lead to so much $PdV$ heating that the fluid element undergoes net heating due to the vibration.  This issue is discussed in \cite{Fernandez:2005cg,Flores-Tulian:2006svb,Wang:2018yux}.  Arras \& Weinberg find that tidal heating from direct Urca processes can heat an inspiraling neutron star to an average core temperature of, at most, 10 keV.  They find little alteration of the inspiral from the tidal excitations.

Hyperonic reactions like $n+p\leftrightarrow \Lambda + p$ have an even faster rate than direct Urca, so they may be able to enhance the tidal heating.  Ghosh, Pradhan, \& Chatterjee \cite{Ghosh:2023vrx} found, with the assumption of tidal oscillations that remain subthermal\footnote{The relationship between the density oscillation magnitude $\delta n_B$ and the deviation from chemical equilibrium $\delta\mu$ depends on susceptibilities of the EoS (see Eq.~\ref{eq:dmu_max_npe} for the relationship in $npe$ matter, which involves the susceptibility $A$) and the speed of the strangeness equilibration $\gamma$ with respect to the density oscillation frequency $\omega$.  I do not know if tidal oscillations would lead to subthermal or suprathermal deviations from strangeness equilibrium.}, that hyperonic reactions can lead to tidal heating of the star to temperatures greater than 100 keV and potentially measurable phase shifts in the inspiral gravitational wave signal.
%%%%%%%%%%%%%%%%%%%%%%%%%%%%%%%%%%%%%%%%%%%%%%
\subsection{Neutron star merger and remnant}\label{sec:postmerger}
What we know about neutron star merger remnant dynamics at this point comes almost entirely from numerical simulations \cite{Baiotti:2016qnr}, as no gravitational wave signal from this phase of the merger has yet been observed due to its high frequency, where the LIGO detector is less sensitive \cite{Baiotti:2019sew}.  The numerical simulations evolve Einstein's equations coupled to relativistic hydrodynamics.  The EoS describing the neutron star matter \cite{Oertel:2016bki} could be a simple analytic parameterization, like a polytrope, or it could be the tabulated result of a nuclear theory calculation (see the CompOSE repository \cite{CompOSECoreTeam:2022ddl}).  Unlike the neutrons, protons, and electrons, which are considered to be a multicomponent fluid whose chemical composition is described by the proton fraction $x_p$, in much of the merger remnant the neutrino MFP is not short enough for the neutrinos to be considered part of the $npe$ fluid.  Intricate transport schemes \cite{Foucart:2022bth} are devised to handle the neutrinos.  

Merger simulations mostly agree upon the following physical picture following the end of the inspiral phase \cite{Baiotti:2016qnr,Baiotti:2019sew}.  When two neutron stars merge, the matter at the interface undergoes dramatic heating due to the compression and from shocks \cite{Perego:2019adq}.  Its temperature may reach many tens of MeV.  The two dense cores retain their structural integrity for some time, bouncing off of each other while gravitational radiation, and perhaps other dissipative mechanisms, damp the energy of the oscillation, bringing them together within ten milliseconds \cite{Most:2021zvc,Chabanov:2023blf}.  As the cores lose their structural integrity, what remains is a differentially rotating mass of dense matter, with a density profile that decreases from core to edge, but with a temperature profile that is peaked at a few kilometers away from the center.  The densest region of the remnant remains cool, likely with a temperature at or below 10 MeV (see Fig.~4 in \cite{particles2010004}).  Neutrinos are continuously being produced by weak interactions in the dense matter, and in areas where the temperature increases beyond several MeV, they become trapped and move with the fluid element.

\begin{figure}
  \centering
  \includegraphics[width=0.5\textwidth]{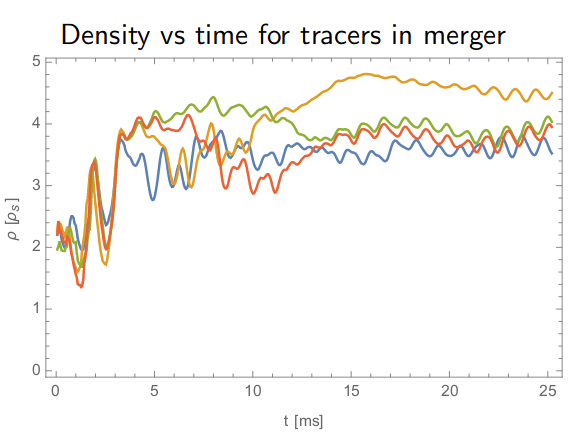}
  \caption{Time evolution of the density of several comoving fluid elements during a neutron star merger.  The tracer particles initially experience large density oscillations right when the two neutron stars touch (t = 0), but after several milliseconds the oscillation amplitude decreases.  The underlying simulation does not include viscosity.  Figure courtesy of M.~Hanauske and the Rezzolla group and originally displayed in \cite{Harris:2020rus}.  A similar figure tracking density \textit{and temperature} of tracer particles is given in Fig.~4 of \cite{Alford:2017rxf}.}
  \label{fig:merger_tracers}
\end{figure}

Tracer particles put into inviscid simulations of the merger remnant indicate the presence of density oscillations on millisecond timescales.  The fluid elements in Fig.~\ref{fig:merger_tracers} start out at $2n_0$ and then experience wild density oscillations when the stars collide, some changing density by over 100\% in the first few milliseconds.  After five ms or so, the remnant has settled down and the fluid element density oscillations become a lot closer to small amplitude, periodic oscillations around an equilibrium density.  The timescale of the density changes matches the collision frequency of the two cores immediately after merger (see Fig.~3 in \cite{Most:2021zvc} and the back-of-the-envelope estimate in footnote \ref{footnote:frequencyestimate}).

Because a neutrino transport scheme already exists in neutron star merger simulations, it makes sense to account for bulk viscosity through the reaction network that gives rise to it, if possible, instead of treating it with a Muller-Israel-Stewart formalism.  These two methods were compared by Camelio \textit{et al.}~\cite{Camelio:2022ljs,Camelio:2022fds} in the case of oscillating neutron stars.  Considering the reaction network approach avoids the need to pre-calculate the transport coefficient $\zeta(n_B,T,\omega)$, which necessarily assumes small amplitude density oscillations (even for the suprathermal case, the  amplitude $\delta n_B\ll n_B$), which Fig.~\ref{fig:merger_tracers} indicates are only found after the first few ms post collision.

\begin{figure*}
  \centering
  \includegraphics[width=\textwidth]{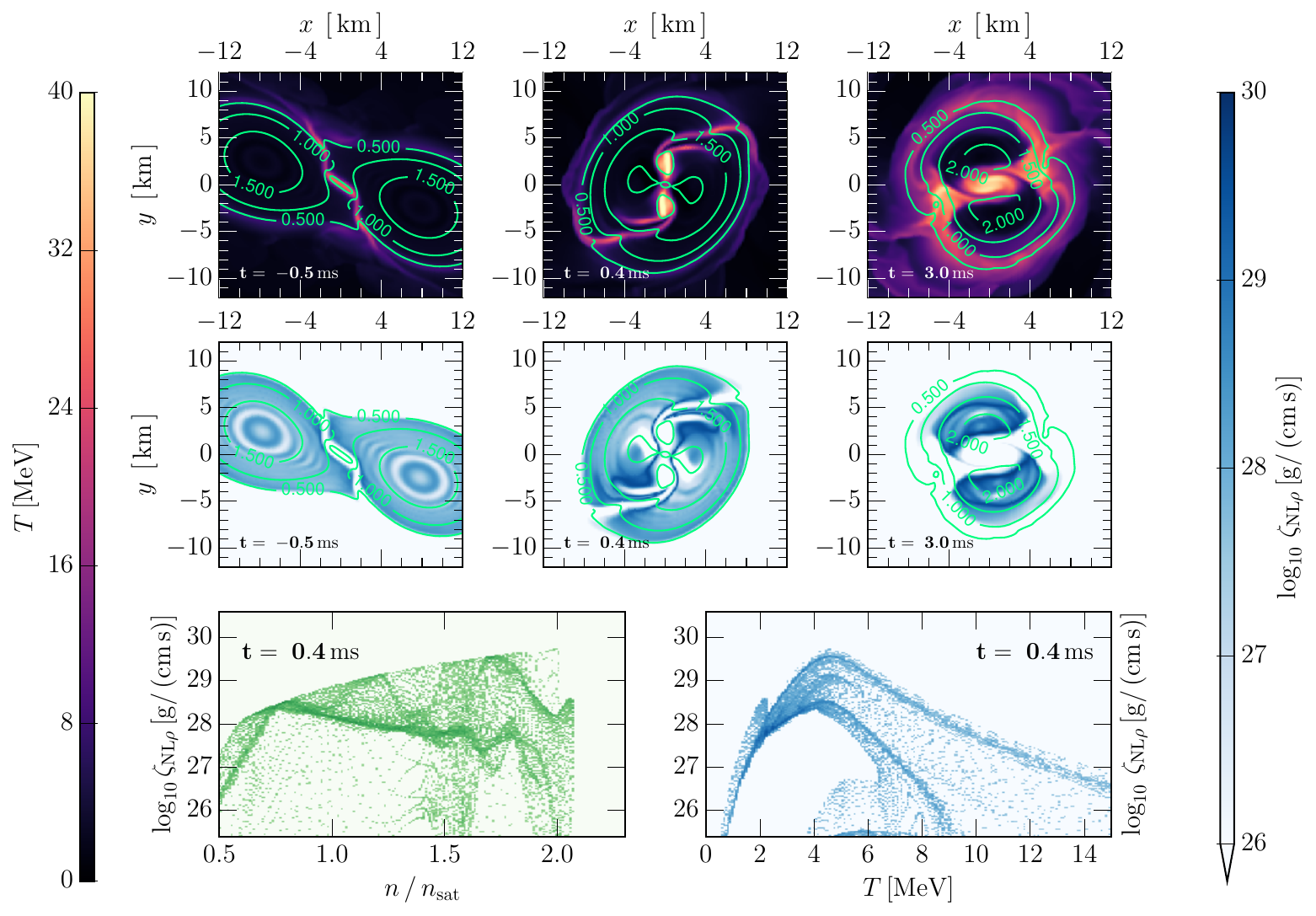}
  \caption{Post-processing of an inviscid neutron star merger simulation to predict the strength of the bulk viscosity.  Top row: The temperature (left color scale) obtained in the equatorial plane of the merger remnant at three time slices.  Center row: The subthermal bulk viscosity (right color scale), assuming a 1kHz density oscillation.  Bottom row: Subthermal bulk viscosity of fluid elements in a merger at one time slice, plotted according to their density (left) and temperature (right).  Figure from Most \textit{et al.}~\cite{Most:2021zvc}.}
  \label{fig:zeta_panel}
\end{figure*}

The data presented in Fig.~\ref{fig:zeta_panel} is from an inviscid neutron star merger simulation \cite{Most:2021zvc}.  The top row shows the temperatures encountered in the merger remnant at different time slices, as well as selected density contours.  In the third panel, the two cores are distinct and are separated by lower density (and hotter) matter.  The middle row depicts the subthermal bulk viscosity calculated according\footnote{The calculation of the equilibration rate $\gamma$ in Ref.~\cite{Most:2021zvc} is slightly different than in the $npe$ section of this chapter.  Direct and modified Urca are both included, as is some artificial blurring of the direct Urca threshold.  See Sec.~2.1 in \cite{Most:2021zvc} for full details.} to Eq.~\ref{eq:bv_npe}, using the density and temperature of each fluid element in the equatorial slice.  The oscillation angular frequency is assumed to be $\omega = 2\pi\text{ kHz}$ and the matter is assumed to be neutrino transparent at all temperatures.  The bottom two panels plot the expected bulk viscosity of each fluid element in one time slice according to the density (left panel) and temperature (right panel) of the fluid element.  The resonant behavior of the subthermal, neutrino-transparent bulk viscosity as a function of temperature is apparent in the bottom right panel.  This post-processed analysis shows that, neglecting back reaction of the bulk viscosity and if the matter remains neutrino transparent, large value of the bulk viscosity are attainable in neutron star mergers.  This figure represents, in some sense, the best case scenario for bulk viscosity in merger remnants, because neutrino-trapping will reduce the bulk viscosity in the high temperature regions of the remnant.  

The implementation of advanced neutrino transport schemes (those that can accommodate out-of-beta-equilibrium physics) in merger simulations is still in its early stages, and simulations are not at the point of being able to identify the impact of weak interactions on energy dissipation.  To study the edge cases, Hammond, Hawke, \& Andersson \cite{Hammond:2022uua} simulated two neutron star mergers, one with infinitely slow beta equilibration and one with infinitely fast beta equilibration.  They found that due to the $x_p$-dependence of the pressure, the gravitational wave signals of these extreme cases are different, thus highlighting the need for simulations to properly take into account weak interactions.  Along the same lines, Most \textit{et al.}~\cite{Most:2022yhe} implemented Urca reactions, calculated in the FS approximation, in a neutrino-transparent simulation and investigated the impact of the weak interactions on the gravitational wave signal.  They found indications that weak interactions alter the gravitational wave signal and generate some entropy.  These simulations are optimistic, as neutrino-trapping would set in at high temperatures and significantly decrease the bulk viscosity in these regions (see Sec.~\ref{sec:nu_trapped}).  

The simulations searching for bulk-viscous effects with the most sophisticated neutrino transport scheme are those of Radice \textit{et al.}~\cite{Radice:2021jtw} and Zappa \textit{et al.}~\cite{Zappa:2022rpd}.  They compare the gravitational wave signals obtained in simulations with advanced neutrino transport schemes that consider chemical equilibration effects with simulations with simpler neutrino transport schemes that do not.  They find that the two schemes produce different gravitational wave signals, but the origin of the difference is still under investigation.  In all of these simulations, the resolution may not be high enough to resolve bulk-viscous effects.  As the bulk viscosity originating from Urca processes peaks at $T\approx 5\text{ MeV}$ \cite{Alford:2023gxq}, which is also around the same temperature at which neutrino trapping becomes important, diminishing the bulk viscosity, it seems clear that further investigation into neutrino transport schemes and out-of-equilibrium effects will be needed to truly understand the role of bulk viscosity in neutron star mergers.

A preview of the potential effects of bulk viscosity on postmerger dynamics was provided in a very recent work by Chabanov \& Rezzolla \cite{Chabanov:2023blf}.  They implemented bulk viscosity with the MIS formalism, and assumed that bulk viscosity took a uniform value, independent of density, temperature, and neutrino-trapping.  If the bulk viscosity is larger\footnote{For context, this condition on the bulk viscosity is only met (in neutrino-transparent $npe$ matter) in a limited temperature window of 3-6 MeV, though across a wide density range (c.f.~\cite{Alford:2023gxq}).} than a few times $10^{29} \text{ g/(cm s)}$, then it can significantly modify the behavior of the merger remnant.  They find that bulk viscosity damps the bouncing of two neutron star cores, as predicted in \cite{Most:2021zvc}.  In addition, the bulk viscosity reduces the bar deformation of the remnant, suppressing the gravitational wave emission and increases the rotational energy of the remnant, raising the $f_2$ peak of the gravitational wave signal to higher frequencies.  
%%%%%%%%%%%%%%%%%%%%%%%%%%%%%%%%%%%%%%%%%%%%%
\section{Conclusions}\label{sec:conclusions}
Bulk viscosity dissipates energy from systems that experience changes in density, provided that they have an internal degree of freedom that is slow to equilibrate.  In dense $npe$ and $npe\mu$ matter, as we have seen, the particle fractions are that degree of freedom, equilibrating through various flavor-changing interactions.   Bulk viscosity in a fluid element undergoing small amplitude oscillations is a function of the oscillation frequency and of the equilibration rates and when the two are comparable, maximal bulk-viscous dissipation occurs.  This stems from the trajectory of the fluid element in the $PV$ plane during one complete oscillation, where if the two rates are equal, then the pressure and volume are out of phase and the $PdV$ work done on the fluid element is maximal.  

Matter in neutron stars undergoes changes in density in many different situations.  A newly born neutron star is likely endowed with oscillations resulting from its violent formation, and the radial oscillations are best damped by bulk viscosity.  Rotating neutron stars generate r-mode oscillations which, at higher temperatures, are kept stable by bulk viscosity.  Inspiraling neutron stars are tidally excited by their companion star and in the course of damping the ensuing oscillations, bulk viscosity can lead to heating of the stars.  Neutron star mergers lead to a violently oscillating merger remnant, and bulk viscosity may be able to dissipate enough energy to alter the gravitational wave signal. 

The use of neutron star mergers to understand transport properties in dense matter, including bulk viscosity, and from a different angle, the need to include effects like bulk viscosity into our simulations of mergers to better predict observables from neutron star mergers, has been the subject of intensive research over the past few years.  If mergers can help us learn about bulk viscosity, this would provide insight into both the weak interaction rates in dense matter and also more obscure properties of the equation of state, like the susceptibilities.  The community realizes the potential for this direction of research, and bulk viscosity in merger environments has been included in recent nuclear theory white papers \cite{Lovato:2022vgq,Achenbach:2023pba} and is part of the science case for Cosmic Explorer \cite{Evans:2023euw,cosmicexplorerpaper}.

On the theory side of bulk viscosity, there is much work to be done.  While it seems likely that bulk viscosity will be included in merger simulations through tracking of the particle degrees of freedom directly (e.g.~Urca reactions in the neutrino transport scheme), calculations of the subthermal and suprathermal bulk viscosity are useful for obtaining an idea of the strength of bulk viscosity as a function of density and temperature.  If the case looks promising, then the effort can be expended to implement the rates in a numerical simulation.  Therefore, it is imperative to calculate the bulk viscosity in other phases of matter, including matter with thermal pions, a pion condensate, quarkyonic matter, and perhaps to revisit quark matter bulk viscosity at higher temperatures.  Even in $npe$ matter there are uncertainties.  The direct Urca threshold is poorly constrained, and the susceptibilities are almost entirely unconstrained.  The peak of bulk viscosity seems to lie at temperatures of around 5 MeV, but at this temperature the assumption of complete neutrino-transparency is likely invalid.  It seems inevitable that improvements in neutrino-transport schemes (especially geared to degenerate matter above densities of $n_0$) will shed light on how bulk-viscous damping occurs as the neutrino MFP shrinks.  

Finally, a wide variety of neutron star merger scenarios should be explored.  Maybe bulk viscosity is most impactful in mergers with certain EoSs, certain mass ratios, or certain orbital eccentricities.  And perhaps other astrophysical events entirely can be additional environments in which bulk viscosity plays a role.  After all, the past few decades have seen the evolution of the focus of bulk viscosity research from the radial oscillations of newly born neutron stars, to the r-modes generated in rotating neutron stars, to the inspiral and merger of two neutron stars.  Maybe bulk viscosity in other scenarios, like white dwarf mergers, white dwarf collapse to a neutron star, or the collapse of neutron stars to twin hybrid star configurations will be the focus of our future attention.
%%%%%%%%%%%%%%%%%%%%%%%%%%%%%%%%%%%%%
\section*{Acknowledgements}
I thank Mark Alford, Alex Haber, and Elias Most for helpful comments on the chapter.  I also want to thank my collaborators on bulk viscosity research: Mark Alford, Bryce Fore, Alex Haber, Elias Most, Jorge Noronha, Sanjay Reddy, and Ziyuan Zhang.  My work is supported by the U.S.~Department of Energy grant DE-FG02-00ER41132.  
\bibliographystyle{JHEP}
\bibliography{references}
\end{document}